\documentclass[floatfix,pra,reprint,amsmath,amssymb,aps,longbibliography]{revtex4-1}
\usepackage{graphicx}
\usepackage{amsfonts}  % Common math fonts
\usepackage{amsmath}   % Common math functions
\usepackage{amssymb}   % Common math symbols
\usepackage{color}
\usepackage{bm}
\usepackage{ascmac}
\usepackage{fancybox}
\usepackage[colorlinks, allcolors = blue]{hyperref}
\usepackage{tabularx}
\usepackage{mathtools}
\usepackage{booktabs}
\usepackage{multirow}
\usepackage[colorinlistoftodos]{todonotes}
\usepackage{soul}

\definecolor{Gray}{rgb}{0.72,0.72,0.98}
\definecolor{LightCyan1}{rgb}{0.83,0.83,0.98}
\definecolor{LightCyan2}{rgb}{0.91,0.92,1}
\begin{document}

% 目次の出力
%\tableofcontents
%\clearpage
%\todo[inline]{The title is to long, perhaps:" Revisiting the exciton binding energy and reduced effective mass in metal tri-halide perovskites"}
\title{Determining Exciton Binding Energy and Reduced Effective Mass in Metal Tri-Halide Perovskites from Optical and Impedance Spectroscopy Measurements}

\author{K. Lizárraga$^{*1,2,3}$}
\author{J. A. Guerra$^{\dagger2}$}
\author{L. A. Enrique-Moran$^{2}$}
\author{E. Serquen$^{2}$}
\author{E. Ventura$^{2}$}
\author{Cesar E. P. Villegas$^{3}$}
\author{A. R. Rocha$^4$}
\author{P. Venezuela$^1$}
\affiliation{$^{1}$Instituto de Física, Universidade Federal Fluminense, Rio de Janeiro, Brazil.}
\affiliation{$^{2}$Departamento de Ciencias, Sección Física, Pontificia Universidad Católica del Perú, Lima 32, Peru.}
\affiliation{$^{3}$Departamento de Ciencias, Universidad Privada del Norte, Lima 15434, Peru.}
\affiliation{$^{4}$Instituto de Física Teórica, UNESP, S$\tilde{\textrm{a}}$o Paulo, Brazil.}

\begin{abstract}  
Accurate determination of the exciton binding energy and reduced effective mass in halide perovskites is of utmost importance for the selective design of optoelectronic devices. Although these properties are currently determined by several spectroscopic techniques, complementary theoretical models are often required to bridge macroscopic and microscopic properties.
Here, we present a novel method to determine these quantities while fully accounting for polarization effects due to carrier interactions with longitudinal optical phonons. Our approach estimates the exciton-polaron binding energy from optical absorption measurements using a recently developed Elliott based Band Fluctuations model.  The reduced effective mass is obtained via the Pollmann–B\"{u}ttner exciton-polaron model, which is based on the Fr$\ddot{\textrm{o}}$hlich polaron framework, where the strength of the electron–phonon interaction arises from changes in the dielectric properties.
The procedure is applied to the family of perovskites ABX$_3$ (A = MA, FA, Cs; B = Pb; X = I, Br, Cl), showing excellent agreement with high field magnetoabsorption and other optical-resolved techniques. The results suggest that the Pollmann–B\"{u}ttner model offers a robust and novel approach for determining the reduced effective mass in metal tri-halide perovskites and other polar materials exhibiting free exciton bands.

\end{abstract}

\maketitle 

    Metal tri-halide perovskites have emerged as a driving force in the field of sustainable energy \cite{hedong}. These materials offer exceptional light absorption and cost-effective production, leading to record-breaking efficiencies of $34.6\%$ in tandem configurations including silicon \cite{longi}. Beyond efficiency, durability improvements have addressed previous stability concerns. Innovations such as robust barrier layers have extended operational lifespans \cite{deokjae}, bringing perovskite cells closer to commercial viability.
    
    To engineer the design and performance of optoelectronic and photovoltaic devices based on these systems, one requires a precise knowledge of the exciton binding energy and reduced effective mass. In particular, the former influences exciton stability, charge carrier diffusion, and light absorption efficiencies \cite{gregg,baranowski}. Meanwhile, the latter dictates carrier mobility limits and modulates the device response to external fields \cite{herz,herzmob,cardona}. Moreover, the reduced effective mass has an effect on the size of exciton radius, the strength of exciton–phonon coupling \cite{baranowski}, the distribution of the density of states near the bandgap, and the electro-optical response in the Franz–Keldysh effect \cite{cardona}.
	
	In metal tri-halide perovskites, the exciton binding energy is commonly determined by optical absorption, photoluminescence (PL), and magneto-optical measurements \cite{baranowski}. These techniques are often combined with theoretical models based on Wannier–Mott excitons. In the case of optical absorption measurements, the dispersion models rely on the  Elliott formulation \cite{elliott,soufiani,sestu,lizarraga}. And, as a result of the different versions of the Elliott formula, one end up with an spread of exciton binding energy values, e.g. $15-45$ meV is reported for MAPbI$_3$ \cite{lizarraga,sestu}. Nevertheless, recent reports points towards low exciton binding energies \cite{galkowski,baranowski}. To shed light in this matter, we calculate the exciton binding energies with a recently developed Elliott based formula that accounts disorder and thermal effects \cite{lizarraga}. 
	
	On the other hand, the reduced carrier effective mass, for metal tri-halide perovskites, is typically estimated by magneto-optical absorption \cite{hirasawa}, high-field magneto-absorption \cite{galkowski}, time-resolved optical spectroscopy \cite{manser,price}, temperature-density resolved optical spectroscopy \cite{yangliu}, and angle-resolved photoemission spectroscopy (ARPES), the latter employed to determinate only the hole effective mass \cite{arpes}. 
    
    \begin{figure*}[!htb] \centering
		\includegraphics[scale=0.4]{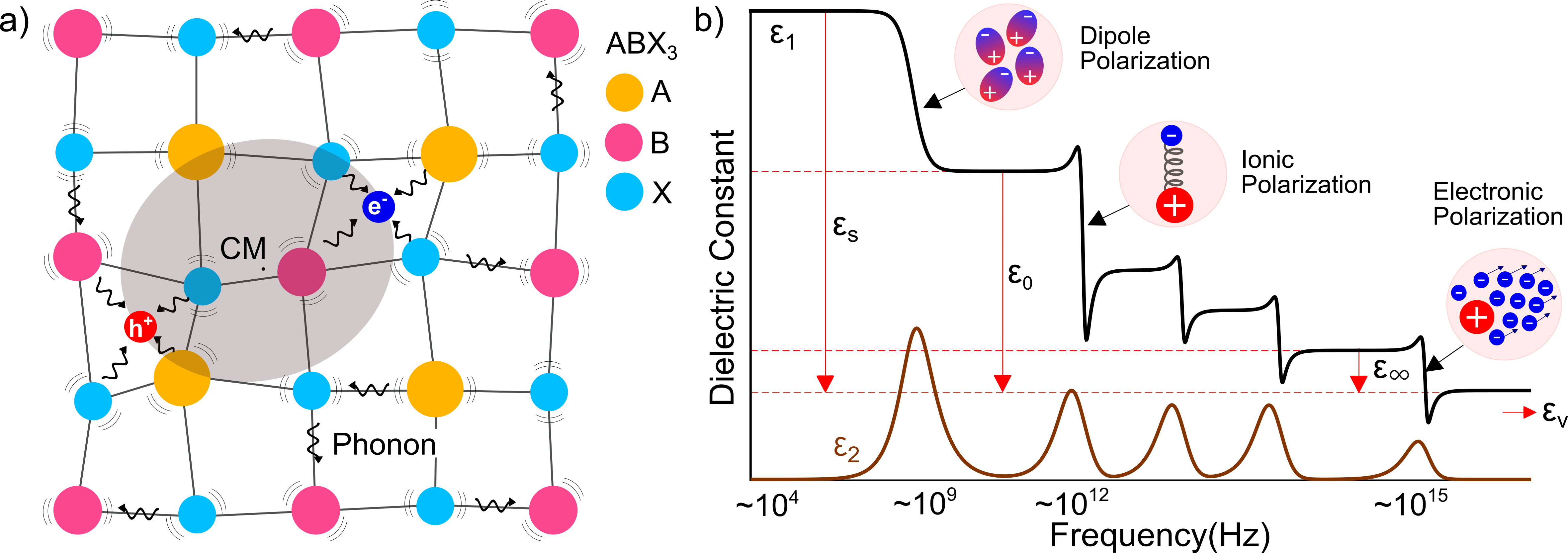}% Here is how to import EPS art
		\caption{(a) Schematic view of the exciton-polaron system as a result of the lattice polarization of a cubic perovskite ABX$_3$ (A: MA, FA,Cs; B: Pb; X: I, Br, Cl). Here, exchange of phonons are represented as zizag arrows and the exciton as the ellipsoid with center of mass at CM. (b) Real ($\varepsilon_1$) and imaginary ($\varepsilon_2$) part of the dielectric function at different frequencies. The dipole, ionic and electronic polarizations are represented as $\varepsilon_s$, $\varepsilon_0$ and $\varepsilon_\infty$, respectively.}
		\label{fig:present} 
	\end{figure*}
	These spectroscopic techniques usually require complementary theoretical models to connect macroscopic and microscopic observables. Indeed, for halide perovskites, these models often disregard polaron formation \cite{baranowski,baranowski2}, even though they are crucial for explaining various phenomena such as low carrier mobility \cite{herzmob}, slow carrier cooling \cite{frost}, and lattice softness \cite{miyatak}. Moreover, polarons renormalize the carrier effective mass, increasing it and leading to photoemission satellites which are described by phonon modes in ARPES measurements \cite{moser}. These polarons can interact with light, forming an exciton–polaron pair. Consequently, lattice polarization screens the Coulomb interaction, resulting in lower exciton binding energies. For instance, $16$ meV for MAPbI$_3$, in contrast to a polaron-less approach which yields energies of the order of $30$ meV \cite{lizarraga,baranowski}. In line with this, the herein presented methodology determines the reduced effective mass accounting polaron effects, showing remarkable agreement.     
	
	Theoretical models describing the exciton‐polaron interactions are many \cite{bajaj,pollmann,kane}. For instance, in the Pollmann‐B\"{u}ttner approach \cite{pollmann}, the exciton binding energy ($E_{xb}$) and the reduced effective mass ($\mu$) are related via the LO phonon energy ($\hbar\omega_{LO}$) and the electronic ($\epsilon_{\infty}$) and ionic ($\epsilon_0$) dielectric constants \cite{pollmann}. In this regard, here we combine the recently developed Elliott-Band-Fluctuations (EBF) optical dispersion model, which has been successfully employed to describe the excitonic properties of paradigmatic semiconductors \cite{lizarraga}, with the Pollmann‐B\"{u}ttner model for exciton‐polaron interactions to accurately determine the exciton binding energy and reduced effective mass of halide perovskites based on MAPbI$_3$, MAPbBr$_3$, MAPbCl$_3$, CsPbBr$_3$, and FAPbI$_3$. To achieve this, we established methods to obtain good estimates of $\epsilon_{\infty}$, $\epsilon_{0}$ and $\hbar\omega_{LO}$, such as spectral ellipsometry measurements in the transparent region, impedance spectroscopy, and the thermal evolution of the exciton linewidth. Finally, we estimate the reduced effective mass from contour maps of the exciton binding energy, and contrast them with previous reports. 
    %This is applied for tri-halide perovskites by taking advantage of the amount of available data in literature.
	%This work presents the absorption model used to obtain the exciton binding energy, the exciton-polaron model, and the methods for estimating the parameters $\epsilon_{\infty}$, $\epsilon_0$, and $\hbar\omega_{LO}$. Section 3 discusses the methodologies for determining these parameters, and presents the results for the exciton binding energy and reduced effective mass. This procedure is  applied to metal tri-halide perovskites based on .
	
	\section{Theory and Methodology} 
	\subsection{Exciton Binding Energy}
    We retrieve the exciton binding energy by means of a recently developed optical dispersion model based on Elliot's equation and the band-fluctuations approach (EBF). The model incorporates potential fluctuations $-$that arise from lattice deformations, disorder and thermally induced localized tail states$-$ of the band edge in the exciton transition rate \cite{lizarraga,guerra}, thus, making it well-suited to depict disorder effects in perovskite systems. The fluctuations are modeled by the distribution function of $\mathcal{G}(z)=e^{z/\sigma}/(\sigma(1+e^{z/\sigma})^2)$, with $\sigma$ being the distribution width, resulting in an analytical expression, containing only six parameters, to describe the bound (discrete) and unbound (continuum) exciton transitions in the absorption coefficient ($\alpha$),  
	\begin{align}
		\alpha&= \frac{A_{1}\sqrt{R^*}}{\hbar\omega} \times \nonumber \\
		&\Bigg( 2R^{*}\sum_{n=1}^{N} \frac{1}{n^{3}\sigma_{n}} 
		\frac{e^{- \left( \hbar\omega-E_{g}+\frac{R^{*}}{n^2} \right)/\sigma_{n}}}{\left( 1+e^{- \left( \hbar\omega-E_{g}+\frac{R^{*}}{n^2} \right) /\sigma_{n}} \right)^2} + \nonumber \\ &A_{2} \left( \frac{1}{1+e^{(E_{g}-\hbar\omega)/\sigma_{c}}}+ \frac{1}{e^{2\pi\sqrt{\frac{R^{*}}{\hbar\omega-E_{g}}}}-1} \right) \Bigg), \label{eq.alphaebf}
	\end{align}
	with,
	\begin{equation}
		\sigma_{n}=\sigma_{c}-\frac{\sigma_{c}-\sigma_{1}}{n^2}, \ \ n=1,2,...,\infty\label{eq.sigma}.
	\end{equation}
	Here, the amplitudes $A_1$ and $A_2$ shape the discrete and continuum components of the $\alpha$ spectrum, and $E_{g}$, $R^{*}$, $\sigma_{n}$, $\sigma_{c}$ correspond to the bandgap, the Rydberg energy, the exciton broadening, and the Urbach energy, respectively. Note that the first exciton binding energy follows the relation $E_{xb}=4R^{*}$. After the best fitted parameters with the EBF model are retrieved, the obtained exciton binding energy is used along with the exciton-polaron model to determine the reduced effective mass.
	
	\subsection{Reduced Effective Mass}
	The reduced effective mass ($\mu$) is determined following the Pollmann‐B\"{u}ttner model, which takes into consideration the exciton-polaron interaction. Below, we outline the model and the main parameters  used in the calculation.
    %($\hbar\omega_{LO}$, $\varepsilon_\infty$ and $\varepsilon_0$)
	
	\subsubsection{Exciton-Polaron Model.}  
	Polarons in halide perovskites are tipically described by Fr$\ddot{\textrm{o}}$hlich theory. In this approach, polarons are formed due to the coupling between electron/hole states and longitudinal optical phonons \cite{Frohlic}. Hence, causing the bare effective mass of electrons ($m_{e}$)  and holes ($m_{h}$) to increase proportionally to the strength of the electron-phonon interaction \cite{Frohlic}.
%	\begin{equation}
%		m_{p}=m_{e/h}\left( 1+\frac{\alpha}{6} \right), \label{alphafrolich}
%	\end{equation}
%	where, $\alpha$ is the Fr$\ddot{\textrm{o}}$hlich coupling constant \cite{Frohlic} defined as:
%	\begin{equation}
%		\alpha=\frac{e^2}{\hbar\varepsilon^{*}}\sqrt{\frac{m_{e/h}}{2\hbar\omega_{LO}}}, \label{eqfrolich}
%	\end{equation}
%	   Eq. (\ref{alphafrolich}) clearly shows the enhancement of the polaron mass due to

	Electron and hole polarons can interact with light, resulting in the creation of exciton-polaron quasiparticles. A schematic figure illustrating the phonon interactions with an exciton is shown in Figure \ref{fig:present}a. The exciton-polaron system has been studied extensively \cite{haken2,bajaj,pollmann,kane} using the effective Hamiltonian for the exciton in a phonon field,
	\begin{equation}
		H_{ex-ph}=\frac{p^2}{2\mu} + E_{s} - V_{eff}(r). \label{eq.e-phham}
	\end{equation}
	Here, the first term is the kinetic energy of the exciton with a reduced mass $\mu=m_e m_h/(m_e+m_h)$ and momentum $p$. $E_{s}$ represents the self-energy of the phonon cloud that surrounds the electron and hole, which is responsible for the tightly bound nature of the exciton‐polaron. $V_{eff}(r)$ is the effective potential of the electron-phonon interaction. Hereafter,  $V_{eff}(r)$ in Eq. (\ref{eq.e-phham}) is replaced by the Pollmann‐B\"{u}ttner (PB) potential \cite{pollmann} due to its ability to produce exciton binding energies comparable with absorption and photoluminescence experiments \cite{baranowski}. This term is a mixture of Coulomb and Yukawa potentials,
	\begin{equation}
		V_{eff}(r) = \frac{e^2}{\varepsilon_{0}r} + \frac{e^2}{\varepsilon^{*}r}\left( \frac{m_h}{\Delta m}e^{-r/l_h} - \frac{m_e}{\Delta m}e^{-r/l_e} \right), \label{eq.pot}
	\end{equation}
	with, $\Delta m=m_h - m_e$. The polaron radius is given by $l_{e/h}=\sqrt{\hbar^2 m_{e/h}/\hbar\omega_{LO}}$. $\hbar\omega_{LO}$ is the LO phonon energy. And, the effective screening $1/\varepsilon^{*} = 1/\epsilon_{\infty} - 1/\epsilon_{0}$, is defined in terms of the ionic ($\varepsilon_{0}$) and electronic ($\varepsilon_{\infty}$) polarizations. The energy range over which the different polarization mechanisms occur in typical polar or ionic materials is depicted in Figure \ref{fig:present}b through the spectral dependence of the complex dielectric function $\tilde{\epsilon} = \epsilon_1+i \epsilon_2$.
	
	The solution of the Schrodinger equation using the Hamiltonian of Eq. (\ref{eq.e-phham}) with the effective potential above, yields the exciton-polaron binding energy. This can be achieved by solving a set of self-consistent equations proposed by Kane \cite{kane} (see eqs. S2-S19 of the supplementary information).
	
	%Here, instead of only considering the s-like displacements of phonon coordinates, as in the PB work, the p-like ones are used as well \cite{kane}. From the latter, a small contribution of $\sim 5\%$ in the exciton binding energy is obtained \cite{kane}. In addition, the implementation of the extr 
	
	The PB potential depicted in Eq. (\ref{eq.pot}) clearly shows that $E_{xb}$ depends on five parameters: $m_{e}$, $m_{h}$, $\hbar\omega_{\text{LO}}$, $\epsilon_{0}$, and $\epsilon_{\infty}$. If the latter three are known from experiments, the PB solution allows one to construct a contour map of the exciton-polaron binding energy as a function of the electron and hole effective masses. The exciton binding energy retrieved from the optical dispersion EBF model (section II-A) is then identified in the exciton-polaron binding energy contour map. Hence, producing a region of possible carrier masses that can be used to estimate the reduced effective mass.
     
	Thus, in order to successfully estimate the reduced effective mass, one requires an accurate determination of both the electronic and ionic dielectric constants, as well as the LO phonon energy. 
    %In the following subsections, we explore the best-practice techniques for retrieving these parameters.
	
	\subsubsection{Electronic Dielectric Constant ($\epsilon_{\infty}$).}
	The electronic (high frequency, or optical) dielectric constant refers to the femtosecond response of electrons, as depicted in Figure \ref{fig:present}b. The method for retrieving this parameter is based on optical spectroscopy, such as spectral ellipsometry or spectrophotometry, covering the energy ranges from the infrared (IR) to visible (VIS) light \cite{fujiwara2}. Retrieving $\varepsilon_1$ and $\varepsilon_2$ typically requires an optical dispersion model to fit measured ellipsometric and/or spectrophotometric measurements, such as phase, amplitud, transmittance, and/or reflectance \cite{fujiwara2,guerra3}. 
    
	The value of $\varepsilon_{\infty}$, defined as the square of the refractive index $n^2(\lambda\rightarrow \infty)$ when $k=0$ \cite{fujiwara2}, can be obtained from the Cauchy optical dispersion model in the transparent spectral region (where $\varepsilon_2 \sim 0$). This region is selected because the effect of phonons and resonant peaks present in the IR and band edge regions can be avoided. The Cauchy equation is a Taylor expansion of the Tauc-Lorentz model for $k=0$. It reads as:
	%\begin{align}
	%	\varepsilon_{1}(\lambda)&=n^{2}(\lambda) = n_\infty^2 + \sum_{i} \frac{C_{i}N_{i}\lambda^2}{\lambda^2 - \lambda_{0i}^2} \nonumber \\ 
    %     \varepsilon_{2}(\lambda)&=0 \label{eq.sellmier}
	%\end{align}
	%Here, $N_i$ and $\lambda_{0i}$ are the amplitudes and the resonant \magenta{wavelengths}, respectively. $C_i$ represents a scale factor.     
    %Similarly, the Cauchy model expresses the dielectric constant as a series expansion of eq. (\ref{eq.sellmier}), i.e.,
	\begin{align}
		n(\lambda)&= n_\infty + 10^2\frac{N_1}{\lambda^2} + 10^7\frac{N_2}{\lambda^4} + ..., \nonumber \\
        k(\lambda)&=0, \label{eq.cauchy}
	\end{align}
	with $n_\infty=\sqrt{\varepsilon_\infty}$, and the successive coefficients $N_{i}$ affecting the curvature and amplitude at shorter wavelengths. In this way, $\varepsilon_\infty$ can be determined by fitting the refractive index in the transparent spectral region. 
    %Eq. (\ref{eq.cauchy}) will be later applied to model the real part of the dielectric constant of the tri-halide perovskites. 

	\begin{figure*}[!htb] \centering
		\includegraphics[scale=0.48]{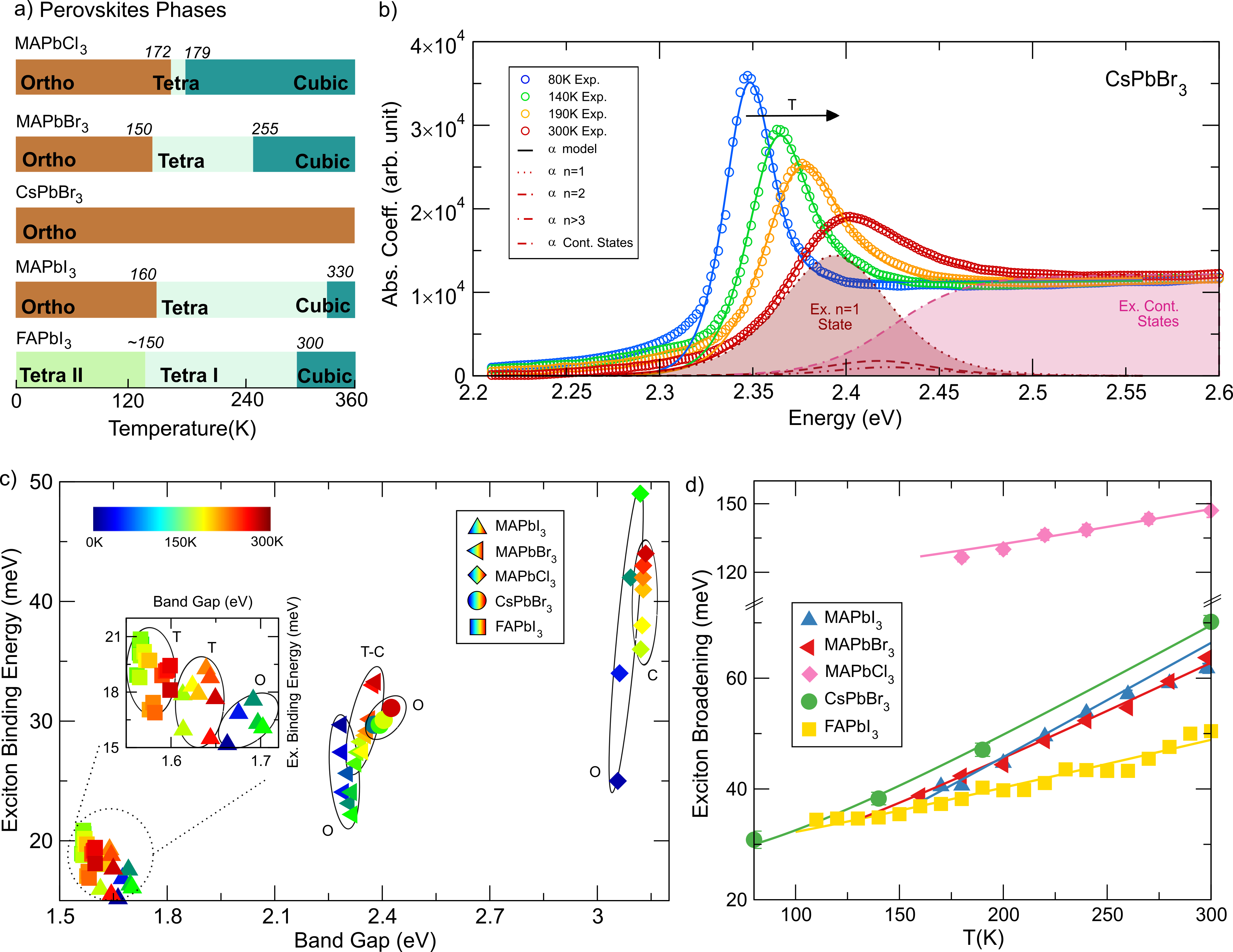}% Here is how to import EPS art
		\caption{(a) Phases of the analyzed tri-halide perovskites as a function of temperature. (b) Temperature-dependent optical absorption data from CsPbBr3, extracted from Wolf et al. \cite{cristoph}, presented as open circles. The solid lines indicate the fitting with the EBF model. Contributions of discrete and continuum exciton states of the EBF model are presented as shaded regions. (c) Exciton binding energies as a function of the bandgap at different temperatures obtained from the EBF model. Note that the corresponding phases of perovskites are enclosed by ellipses and labeled as orthorhombic (O), tetragonal (T) and cubic (C). The inset present a zoom around 1.6eV for MAPbI$_3$ and FaPbI$_3$.(d) Evolution of the exciton broadening (FWHM) for temperatures in the range of the analyzed phases. The solid lines represent the fitting of Eq. (\ref{eq.linewidth}) to obtain the LO phonon energy.}
		\label{fig:parameters} 
	\end{figure*}
	
	\subsubsection{Ionic Dielectric Constant ($\epsilon_{0}$).}
	The picosecond response of the lattice vibrations due to phonons defines the ionic (low-frequency) dielectric constant (see Figure \ref{fig:present}b). For metal tri-halide perovskites, this parameter is studied with optical- (in the IR spectral region) and impedance- spectroscopy \cite{stuart,wilson}. While optical measurements are biased by energy loss mechanisms within the medium, known as dielectric relaxation losses, impedance spectroscopy is a technique capable of accounting for them \cite{wilson,barsoukov}. These measurements, when carried out over a wide range of frequencies, exhibit a decrease for the real part of the dielectric constant as well as a resonant peak for the imaginary part when the system is polarized \cite{blythe}, as depicted in Figure \ref{fig:present}b. This damping can be modeled using the Debye relaxation model, $\varepsilon_1=\varepsilon_0 + \frac{\varepsilon_s-\varepsilon_0}{1+\omega^2\tau^2}$, where $\tau$ is the relaxation time of dipoles, $\varepsilon_s$ and $\varepsilon_0$ are the dipole and ionic dielectric constants, respectively \cite{polardebye,poglitsch}. In particular, for halide perovskite systems, extrinsic contributions such as grain boundaries affect electronic relaxation process \cite{govinda}. These contributions are typically modeled by adding the Maxwell-Wagner term to the Debye relaxation model \cite{govinda}. Thus, the real part of the dielectric constant is described as:
    \begin{align}
		\varepsilon_1=\varepsilon_0 + &\frac{\varepsilon_s-\varepsilon_0}{1+\omega^2\tau^2} + \varepsilon_0^{ext}+\frac{\varepsilon_s^{ext}-\varepsilon_{0}^{ext}}{2} \nonumber \\
        &\times \left( 1-\frac{\textrm{sinh}\left[ \beta \ \textrm{ln}(\omega\tau^{ext}) \right] }{ \textrm{cosh}\left[ \beta \ \textrm{ln}(\omega\tau^{ext}) + \textrm{cos}(\beta\pi/2)  \right] } \right). \label{eq.debye2}    
	\end{align}
	Here, $\beta$ is a dimensionless constant varying from 0 up to 1, $\varepsilon_s^{ext}$ and $\varepsilon_0^{ext}$ are the dipole and ionic dielectric constants from the external factors, respectively. Additionally, the thermal energy of interacting dipoles lead to a temperature-dependent difference of dielectric constants $\varepsilon_s-\varepsilon_0=C/(T-T^{*})$ \cite{govinda,onoda2}, where, $T^{*}$ is a constant accounting the dipolar interactions, and $C=n\tilde u^2\eta/3Vk_\beta \varepsilon_v$ is a system dependent constant porportional to the number ($n$) and strength ($\tilde u$) of dipole moments, lattice local field ($\eta$) and inversely dependent on the volume of unit cell ($V$) \cite{govinda}. Therefore, we suggest that the determination of $\varepsilon_0$ should be carried by impedance spectroscopy modeled with the Debye relaxation including polarization layers.
	%Remarkably, the terms $\varepsilon_s$ and $\varepsilon_0$ in eq. (\ref{eq.debye}) could change depending on the spectral region of the measurements. For instance, the former changes to $\varepsilon_0$, while the latter to $\varepsilon_\infty$ when the analyzed region covers the ionic to electronic response. When dipole region is analyzed, as in eq. (\ref{eq.debye})

    % where the relaxation time of the dipoles is expressed as $\tau(T)=1/(T-T_{0})$, or alternatively, $\tau=\tau_{\infty}e^{A/T}$ \cite{anusca,poglitsch}. 
    
	\subsubsection{LO Phonon Energy ($\hbar\omega_{LO}$).}
	Methods for estimating the LO phonon energy are based on the thermal evolution of the optical bandgap and the temperature dependence of the Full Width at Half Maximum (FWHM) of the exciton absorption band. 

   On the one hand, for perovskite systems, the thermal evolution of the bandgap hinders the correct determination of the LO phonon energy because of the predominant contribution from the volumetric thermal expansion \cite{wei,francisco2019}. On the other hand, the thermal evolution of the exciton band FWHM can be expressed as the sum of contributions arising from the interactions of electrons with both acoustic and optical phonons \cite{segall,rudin}, with the latter being shaped by the Bose-Einstein distribution for phonon occupancies \cite{rudin,wei}, 
	\begin{equation}
		\Gamma(T)=\Gamma_{0}+\Gamma_{AC}T+\Gamma_{LO} \left( \frac{1}{e^{\hbar\omega_{LO}/k_{\beta}T}-1} \right). \label{eq.linewidth}
	\end{equation}
	Here $\Gamma_{AC}$ and $\Gamma_{LO}$ are proportionality constants related to acoustic and optical phonon terms, respectively \cite{wei}. Nevertheless, the electric field produced by the longitudinal phonon modes is stronger than transverse ones in polar materials \cite{kittel}. Thus, allowing neglect AC phonons and take only LO phonons into account when describing perovskite systems \cite{adam,bernhard}. Based on the description of the FWHM, we believe that temperature-dependent ellipsometry measurements would be a good approach to estimate $\hbar\omega_{LO}$.
	
	\section{Results and Discussion}
    In the following sections, we determine the exciton binding energy from the EBF optical dispersion model, the exciton-polaron parameters from optical- and impedance- spectroscopy, and the reduced effective mass from the surface maps of exciton binding energy. 
    
	\subsection{Exciton Binding Energy}
	Temperature-dependent optical absorption measurements in this work were retrieved from Soufiani et al. for MAPbI$_{3}$ and MAPbBr$_3$ \cite{soufiani,lizarraga}, Saxena et al. for MAPbCl$_3$ \cite{saxena}, Wolf et al for CsPbBr$_3$ \cite{cristoph}, and Xu et al. for FAPbI$_3$ \cite{youcheng}. These were fitted using the EBF model of Eq. (\ref{eq.alphaebf}). Figure \ref{fig:parameters}a-b depict the different phases of perovskites in the analyzed temperature range of $0-360$K and the absorption coefficient fitted curves for CsPbBr$_3$, respectively. The latter showing the characteristic blue shift of the absorption when temperature raises, along with the contributions from the n$^{th}$ discrete and continuum exciton states. For the remaining systems, the fitting curves and best fitted parameters are presented in the supplementary information (SI).
	
	Figure \ref{fig:parameters}c shows the behaviour of the exciton binding energy as a function of the bandgap for various perovskites and temperatures. Here, we observe an increase in the exciton binding energy when the halide ion changes from I$\rightarrow$Br$\rightarrow$Cl. This trend is a consequence of the decrease in the electronic response ($\varepsilon_\infty$), which, as we will show, drops from $5.0\rightarrow3.8\rightarrow3.3$. Moreover, variations in the cation A (MA, FA, Cs) appear to have no significant impact on either the exciton binding energy or the bandgap. This behavior can be attributed to the similar lattice parameters and electronic polarizabilities found in these perovskite families \cite{bokdam}, as exemplified by MAPbBr$_3$ and CsPbBr$_3$ (see Table I).
	
	The well-known phase transition in perovskites is reflected in the behavior of the bandgap in figure \ref{fig:parameters}c. For instance, in the case of MAPbI$_3$, the bandgap shows two distinct clusters above and below 160$K$, as a consequence of the phase transition from orthorhombic to tetragonal phases\cite{leguy2}. For MAPbBr$_3$ and MAPbCl$_3$ such transition takes places at $\sim$150$K$ and $\sim$179$K$, albeit with a smaller effect on $E_g$. Note that CsPbBr$_3$ and FAPbI$_3$ do not exhibit these effects since their crystal line structures are only slightly modified: the former remains orthorhombic \cite{svirskas}, while the latter undergoes a small transition from tetragonal I to tetragonal II \cite{junwen,tianran}, as depicted in Figure \ref{fig:parameters}a.
	%\todo[inline]{The paragraph above was not clear at all. Take as an example MAPbI3, it is say "The bandgap shows a discontinuity" but talk about a discontinuity implies a defined function suddenly drops to a particular value. What I see here is that from 0 to 160K the binding energy increases and then slightly drops around ~130K. Then the binding energy for temperatures T $>$ 160K posses lower bandgaps and behaves without a clear trend, i.e, goes up then drops then goes down. Check out my suggestions} 
    
	The exciton binding energy at 300K are summarized in Table \ref{table:parameters}. Our low values of the exciton binding energy are in good agreement with the recently reported values based on other experimental techniques \cite{yakovlev,galkowski,zang,sun,niesner}. Hence, suggesting the capability to obtain the same accuracy of other more-demanding techniques, such as high-field magnetoabsorption, by modeling of the optical absorption with the EBF model.
    %based on optical absorption analysis \cite{yangliu,jeon,fabian,comin,soufiani}, magneto-absorption \cite{yakovlev}, high-field magneto-absorption \cite{galkowski,zang}, and photoluminescence \cite{sun,niesner}.
    %\starr{In the case of MAPbI$_3$ and FAPbI$_3$, an exciton binding energy of $\sim18$~meV is obtained. For MAPbBr$_3$ and CsPbBr$_3$, a value of $\sim32$~meV is estimated, while for MAPbCl$_3$, a larger value of $44$~meV is retrieved.} \todo[inline]{not necessary as it is already in the table}

    \begin{table*}
		\centering
		\begin{tabular}{ m{2.0cm}|  m{1.2cm} m{2.0cm}  m{1.0cm} m{2.0cm} m{2cm} m{1.2cm}}
			\hline\hline 
			Material & Phase & $E_{xb}$(meV) & $\varepsilon_{\infty}$ & $\varepsilon_{0}$ & $\hbar\omega_{LO}$(meV)  &$\mu$($m_e$)\\
			\hline
			$\textrm{MAPbI}_{3}$ & T &  16.3$\pm 1.6$ & 5.0 & 22.0 \cite{govinda} & 10.0$\pm4.1$  & 0.107\\
            $\textrm{MAPbBr}_{3}$ & C &  33.3$\pm0.5$ & 3.8 & 20.0 \cite{govinda} & 17.8$\pm5.1$  & 0.121\\
            $\textrm{MAPbCl}_{3}$ & C &  44.0$\pm6.0$ & 3.3 & 22.1 \cite{onoda2} & 26.2$\pm2.1$  & 0.135\\
			$\textrm{CsPbBr}_{3}$ & O  &  31.1$\pm1.5$ & 3.9 & 20.5 \cite{govinda}  & 19.5$\pm8.6$ & 0.128 \\ 
			$\textrm{FAPbI}_{3}$ & T & 18.1 $\pm0.2$ & 4.5$^{C}$ & 17.7 \cite{caselli}  & 14.4 $\pm6.0$  & 0.103\\
			\hline\hline
		\end{tabular}
		\caption{Summary of results. $\varepsilon_{\infty}$, Electronic dielectric constant retrieved after the Cauchy fitting for the transparent region of ellipsometry measurements from literature reports. $\varepsilon_0$, Ionic dielectric constant extracted from impedance spectroscopy measurements. $\hbar\omega_{LO}$, LO phonon energy retrieved after fitting Eq. (\ref{eq.linewidth}) to the thermal evolution of the exciton broadening. $E_{xb}$, Exciton binding energy at 300K calculated from the fitting of Eq. (\ref{eq.alphaebf}) to the optical absorption. And, average reduced effective mass, $\mu$, determined in this work. Noteworthy that the here presented values correspond to the phase of the perovskite at 300K. Phases at 300K are denoted as tetragonal (T), cubic (C) and  orthorhombic (O). Only in the case of FAPbI$_3$, the value of $\varepsilon_{\infty}$ for the cubic phase was selected. \label{table:parameters}}
	\end{table*}
    
	\subsection{Exciton-Polaron Parameters}

	\subsubsection{Electronic Dielectric Constant ($\varepsilon_\infty$).}
	Ellipsometry measurements of MAPbI$_3$ in the UV-VIS range have been carried out in previous studies using different approaches to estimate the dielectric constant \cite{guerra3,isabel,jiang,leguy,loper,shirayama}. These studies suggest a spread of $\varepsilon_\infty$ values (e.g. $4.0-5.6$ is found for MAPbI$_3$). We argue that this apparent discrepancy can be resolved by using Cauchy’s formula (Eq. (\ref{eq.cauchy})) to model the transparent region. That is because, within the models used in those works, the Critical Point (CP) model tends to overestimate the extinction coefficient in the transparent spectral region \cite{shirayama}. Similarly, the Forouhi-Bloomer model is not Kramers-Kronig consistent, so it fails to predict asymptotic behaviors \cite{fujiwara}, leading to misrepresented values of $\varepsilon_\infty$. Therefore, we have extracted the data from the aforementioned reports and fitted it according to Eq. (\ref{eq.cauchy}). The fitting results are presented in Figure S4 and Table S4 of the SI. 
    
    In case of MAPbI$_3$, in the tetragonal phase, our analysis leads to $\varepsilon_\infty$ values between $4.9$ and $5.2$. These are in agreement with the results of Jiang et al. using the Cauchy's model \cite{jiang}, and with the report of Shirayama et al. using Tauc-Lorentz (TL) oscillators \cite{shirayama}. The latter agreement is a consequence of the asymptotic behavior of the TL model in the transparent region (see section B.2). Notably, in Guerra et al. a value of $5.1$ is reported after using a point-by-point method to avoid the use of optical dispersion models \cite{guerra3}. Note that the low spread of our recalculated $\varepsilon_\infty$ does not affect the determination of the exciton polaron binding energy. For this reason, we have chosen the averages summarized in Table \ref{table:parameters}. 
    %This method requires multiple measurements at different angles to add sufficient redundancy to be able to retrieve the optical constants at each measured wavelength. 
    
    The electronic dielectric constant strongly depends on the halide ion, as evidenced by the decrease in $\varepsilon_\infty$ from $5.0$ for MAPbI$_3$ to $3.8$ for MAPbBr$_3$ and $3.3$ for MAPbCl$_3$. Notably, CsPbBr$_3$ and FAPbI$_3$ perovskites show similar values to those of the compounds based on MA with bromide and iodide, respectively. Finally, in the case of FAPbI$_3$, the  value of $\varepsilon_\infty$ for the cubic phase was selected due to a lack of reports on the tetragonal phase. The similarities in the optical response between both phases support this choice \cite{muhammad}.
    %In the case of CsPbBr$_3$, a reduction in the dielectric response (with $\varepsilon_\infty=2.9$) is observed when the cubic phase is formed \cite{minglinzhao,washeng}.
	
	In essence, the proposed method, which leads to a reduction in the accessible values of $\varepsilon_\infty$, decreases the spread of values for the exciton-polaron binding energy \cite{menendez}. This consensus values seem to be in agreement with some theoretical reports based on density functional perturbation theory (DFPT)\cite{du,menendez,welch,sapori}. 
    %For instance, values of $5.1-5.3$, $4.2$, and $3.8$ are obtained for MAPbI$_3$, MAPbBr$_3$, and CsPbBr$_3$, respectively.
	
	\subsubsection{Ionic Dielectric Constant ($\varepsilon_0$).}
%    On the opposite, if defect layers are ignored, one can obtain larger values of $28.8$ or $30.0$ \cite{anusca,poglitsch}. Interestingly, low values of $23.3$ can be also obtained by modeling the permittivity as that of a polar liquid using the Kirkwood-Fr$\ddot{\textrm{o}}$lich equation \cite{onoda}. 
	We now turn our attention to the determination of the ionic dielectric constant. Table \ref{table:parameters} summarizes the values obtained from the literature for halide perovskites in the corresponding phase at room temperature. For MAPbI$_3$, this corresponds to $\varepsilon_0=22.0$, which arises from the fitting of Eq. (\ref{eq.debye2}) in the dielectric constant obtained by impedance spectroscopy on powder samples of the tetragonal structure \cite{govinda}. Notably, theoretical studies corroborate this choice \cite{du,brivio}. A similar analysis can be performed for MAPbBr$_3$, for which the same methodology report a value of $\varepsilon_0=20.0$ in the tetragonal phase \cite{govinda}. The latter value can be extended to the cubic phase arguing that both phases share similar lattice parameters \cite{lehmann,leguy2}.
    
    Nevertheless, to the best of our knowledge there are no further reports of the Debye model including extrinsic contributions for cubic MAPbCl$_3$. Thus, leading us to select the value of $\varepsilon_0=22.1$  from Onoda et al. \cite{onoda2}. On the other hand, in the case of orthorhombic CsPbBr$_3$, the work of Govinda et al. suggests that $\varepsilon_0=20.5$ remains constant over a wide range of frequencies and temperatures \cite{govinda}. Finally, for tetragonal FAPbI$_3$, Caselli et al. reported a value of $\varepsilon_0=17.7$ at 1.7 MHz \cite{caselli}.
    %due to the absence, to the best of our knowledge, of reports on the cubic phase analyzed using defect polarization layers. 
	
	\begin{figure*}[!htb] \centering
		\includegraphics[scale=0.43]{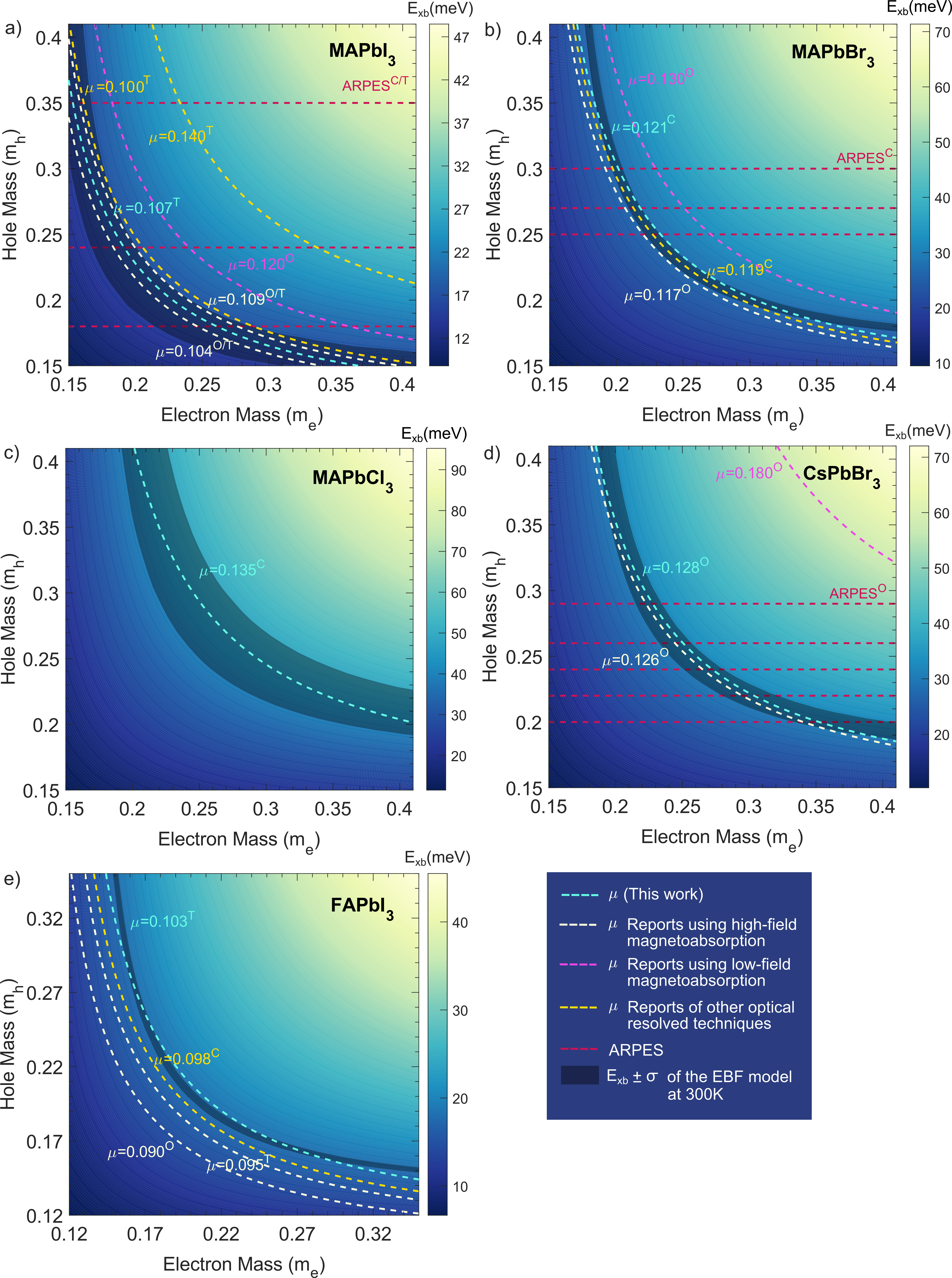}% Here is how to import EPS art
		\caption{Exciton-polaron binding energy surface map, calculated with Pollmann's model, as a function of the electron (x-axes) and hole (y-axes) bare effective masses for the $\textrm{MAPbI}_{3}$ (a), $\textrm{MAPbBr}_{3}$ (b), MAPbCl$_3$ (c), $\textrm{CsPbBr}_{3}$ (d) and $\textrm{FAPbI}_{3}$ (e). Here, the exciton binding energy shown in Table \ref{table:parameters} is presented as the black shaded region. Note that the average reduced effective mass is presented in sky-blue lines. Reduced effective masses from other experimental reports are presented in yellow dashed lines where the superscript denotes the perovskite phase, orthorhombic (O), tetragonal (T) and cubic (C). The hole effective mass retrieved from ARPES measurements are presented as magenta horizontal lines. Note that the carrier effective masses are in terms of the electron mass}
		\label{fig:massperovs} 
	\end{figure*}	
    
	\subsubsection{LO phonon Energy ($\hbar\omega_{LO}$)}
    
	To determine the LO phonon energy, we model the thermal evolution of the exciton broadening, obtained from the EBF model, with Eq. (\ref{eq.linewidth}). This procedure is applied to the temperature ranges of the halide perovskites at the principal (room temperature) phase to avoid phase transition effects. For instance, MAPbI$_3$ in the tetragonal phase covers temperatures of $160-330$K, and MAPbCl$_3$ in the cubic phases covers the range from 180 to 300K (see Figure \ref{fig:parameters}a). Note that, in the case of MAPbBr$_3$, we have selected data points from the tetragonal and cubic phases, arguing that both phases share similar lattice parameters \cite{lehmann}. 
    
    The fitting procedure of Eq. (\ref{eq.linewidth}) along with the best fitted parameters can be found in section S3 of the SI. The results for the halide perovskite systems are presented in Figure \ref{fig:parameters}d, and the LO phonon energies are summarized in Table \ref{table:parameters}. Overall, these results are in agreement with other reports extracting the LO phonon energy from thermal evolution of absorption and photoluminescence measurements \cite{lizarraga,adam,saxena,sendner,cristoph,iaru,mingfu}. Moreover, the here reported values are also within the highest LO phonon modes (Cage modes) obtained from Raman spectroscopy for MA- \cite{leguy2}, Cs- \cite{keisei}, and FA- \cite{steele} based perovskites.
    
%	For instance,     
%    Nonetheless, the different behavior of the exciton broadening data points (see Figure \ref{fig:parameters}d) leads to a misleading estimation $\hbar\omega_{LO}$. For instance, MAPbI$_3$ and MAPbBr$_3$ in the low-temperature regime (orthorhombic phase) exhibit a flat trend in exciton broadening, in contrast to their increasing behavior in the tetragonal or cubic phases. Similarly, MAPbCl$_3$ presents a discontinuity at $~170$K, caused by the phase transition from orthorhombic to cubic.     
%    Thus, in order to neglect the impact of the phase transition, we have selected the exciton broadening data points at a given phase.
	\subsection{Reduced Effective Mass}
	
	As described in section I-B, the PB model can be used to estimate the reduced effective mass given prior knowledge of the electronic ($\varepsilon_\infty$) and ionic ($\varepsilon_0$) dielectric constants, the LO phonon energy ($\hbar\omega_{LO}$), and the exciton binding energy ($E_{xb}$). This can be achieved as follows. First, the exciton binding energy is computed over a range of carrier effective masses. The resulting surface map of $E_{xb}$ is presented in Figures \ref{fig:massperovs}a–e. Then, the exciton binding energy values retrieved from the optical absorption measurements are selected. Their corresponding error bars are indicated by the black shaded region in Figures~\ref{fig:massperovs}a–e. Lastly, an effective reduced mass is determined from the average of the available carrier effective masses. These are highlighted as sky-blue dashed lines in Figures \ref{fig:massperovs}a–e and summarized in Table \ref{table:parameters}. Note that the value of $E_{xb}$ was selected at 300 K to be consistent with the temperature settings of the selected ellipsometry and impedance measurements, which retrieve $\varepsilon_\infty$ and $\varepsilon_0$, respectively. 
    
%Nonetheless, the calculated value of 0.121 compares well with 0.130 of FAPbBr$_3$ \cite{galkowski}. 
    The consistency of our results with high-field magnetoabsorption measurements (white dashed lines in Figure \ref{fig:massperovs}) arises from the suppression of polaron formation when strong magnetic fields are applied \cite{galkowski,soufiani3,miyata,zang}. In such conditions, the high cyclotron velocities of charge carriers prevent the formation of lattice distortions due to the lattice's inability to quickly respond. \cite{baranowski2}. In contrast, in the low magnetic field regime (dashed pink lines in Figure \ref{fig:massperovs}), the LO phonon energy exceeds the cyclotron energy. This allows enough time for carriers to interact with the lattice, resulting in the formation of polarons, and leading to larger effective masses \cite{tanaka,hirasawa,yakovlev}. Notably, recently the exciton-polaron effects were included into the theoretical model used in low-field magnetoabsorption, resulting in effective masses comparable to those obtained in the high-field limit \cite{baranowski2}. This observation aligns well with our approach, which explicitly accounts for exciton-polaron contributions for retrieving the reduced effective mass. In contrast, methodologies based solely on optical spectroscopic techniques often neglect these effects \cite{yangliu,price,andrew}. 
    
    Although high field magnetoabsorption yields consistent values with those obtained in this work, our method offer a simpler approach that is not limited by  instrumentation equipped with high field magnets or low temperature set ups. The agreement even extends to the cubic MAPbBr$_3$, where the low temperatures used in magnetoabsorption have only explored the orthorhombic phase \cite{galkowski}. Figures \ref{fig:massperovs}a–e also show ARPES results, offering a visual guide for estimating electron effective masses based on known hole masses, for example in MAPbI$_3$, a hole mass of $0.24$ \cite{jinpeng}, corresponds to an electron mass of $0.19$.   
%    The reason behind our concordance with high field magnetoabsorption is based on the fact that the formation of polarons is prevented when high fields are reached. This due to the high magnetic fields producing fast cyclotron velocities of the carrier that the lattice deformations can not follow \cite{baranowski2}. On the contrary, in the low magnetic field limit (highlighted in pink in figure \ref{fig:massperovs}), the LO phonon energy is larger than the cyclotron frequency, thus, properties are affected by polaron formations resulting in large effective masses \cite{tanaka,hirasawa,yakovlev}. Interestingly, if one consider the exciton-polaron theory in the description of low field magnetoabsorption, the reduced effective mass agrees well with the high-field limit \cite{baranowski2}. Remarkably, this align with our method that determines the reduced effective mass by leaving in consideration the exciton-polaron effects. On the other hand, methods based on pure optical-resolved techniques (See Tables S8 and S9 for further information) mostly disregard this effects \cite{yangliu,price,andrew}. Figures \ref{fig:massperovs}a-e also present ARPES results for serving as a guide for the eye to estimate the electron effective mass given the hole mass, e.g. for MAPbI$_3$, a hole mass of $0.24$ \cite{jinpeng}, retrieves an electron mass of $0.19$.
    
    In the case of MA-based perovskites, our results point towards an increase in the reduced effective mass (0.107$\rightarrow$0.121$\rightarrow$0.135) when the halide ion is substituted with lighter ones, i.e., I$\rightarrow$Br$\rightarrow$Cl. This can be explained in terms of the substitution impact on the structural and electronic properties. In perovskites, the lattice parameter is reduced when the halide ion decreases in size \cite{ungi}. A smaller lattice constant results in a stronger overlap between the orbitals arising from the lead and halide ions, which shapes the conduction and valence states, respectively \cite{monika}. This reduced band curvature generates higher effective masses.
%from iodide to bromide to chloride

    The iodide- (MAPbI$_3$ and FAPbI$_3$) and bromide‐based (MAPbBr$_3$ and CsPbBr$_3$) perovskite families seem to have similar reduced effective masses of $0.107\sim0.103$ and $0.121\sim0.128$, respectively, in agreement with the findings of previous reports \cite{galkowski,yangliu}. The latter behavior can be attributed to the similar exciton binding energies and dielectric screening present in these materials. Notably, studies by Herz et al. showed that substituting the Cs atom with an MA/FA cation does not dramatically affect the optical and transport properties,  suggesting that such properties depend only on the inorganic cage, e.g. PbI$_3$ \cite{herzmob,herz}. 
    
    Furthermore, the low reduced effective masses of the systems based on MAPbI$_3$ and FAPbI$_3$ are comparable to those of inorganic semiconductors such as GaAs ($\mu\approx0.071$). As such, a high mobility on the order of $\sim1000$ cm$^2$/(Vs) could be expected \cite{nathan}. However, the measured mobilities, on the order of $<100$ cm$^2$/(Vs), result from the low energies of LO phonons ($\sim10-14$ meV) and the strength of the Fr$\ddot{\textrm{o}}$hlich interaction ($\alpha$) \cite{herzmob}. This is reflected in the Fr$\ddot{\textrm{o}}$hlich coupling constant in Table \ref{table:frolich}, for which perovskites based on bromide and chloride exhibit the largest coupling.
	
	\begin{table}[h]
		\begin{center}
			\begin{tabular}{ m{2.0cm}|  c }
				\hline\hline
				Material & $\alpha_{e-ph}=\frac{e^2}{\hbar\varepsilon^{*}}\sqrt{\frac{\mu}{2\hbar\omega_{LO}}} $\\
				\hline
				MAPbI$_3$& 1.87 \\
				MAPbBr$_3$& 2.05 \\
				MAPbCl$_3$& 2.16 \\
				CsPbBr$_3$& 1.96 \\
				FAPbBr$_3$& 1.64
                
                \\
				\hline\hline
			\end{tabular}
		\end{center}
		\caption{\label{table:frolich} Fr$\ddot{\textrm{o}}$hlich coupling constant ($\alpha_{e-ph}$) calculated with the values from Table \ref{table:parameters}.}
	\end{table}
	
	Finally, we compare the results obtained from our analysis to those obtained from first-principles calculations, finding overall good agreement for the reduced effective masses (See Tables S7). The best agreement is observed with results obtained via methods that predict a correct bandgap \cite{sajedi,baiqing}. These include those obtained with the hybrid HSE functional in DFT and the G$_0$W$_0$ many-body perturbation theory, with the inclusion of spin-orbit coupling and thermal fluctuations \cite{bah,bokdam,becker,sanjun,baiqing}. Similarly, the inclusion of quasi-particle corrections leads to better agreements with ARPES measurements \cite{sajedi}. 
		
	\section{Summary and Conclusions}
	
In this work, we provided a systematic study of the exciton binding energy and the reduced effective mass in organic (FA/MA) and inorganic (Cs) lead (Pb) tri-halide (I/Br/Cl) perovskites. Exciton binding energies were obtained by fitting the absorption coefficient considering the optical dispersion Elliott-Band-Fluctuations model, which takes into consideration disorder effects of the band-edge. The reduced effective masses were retrieved after comparing the surface plots of exciton-polaron binding energy from the Pollmann-B\"{u}ttner model with the one obtained from the dispersion models. To obtain an accurate description of the exciton-polaron system, we have establish the methods to determine the electronic ($\varepsilon_\infty$), and ionic ($\varepsilon_0$) dielectric constants as well as the LO phonon energy ($\hbar\omega_{\text{LO}}$), from experiments that are relatively easy to perform. We found that, for determining $\varepsilon_\infty$, ellipsometry measurements in the transparent region of $\varepsilon_1$, modeled through Cauchy’s (or similar) model, are desirable. Moreover, we argue that the preferred technique for estimating $\varepsilon_0$ is impedance spectroscopy combined with the Debye relaxation model including the Maxwell–Wagner term for accounting extrinsic contributions. Our analysis also suggests that for assessing $\hbar\omega_{LO}$, modeling the exciton broadening in temperature-dependent optical absorption measurements provides a good estimate. Based on these considerations, we obtained exciton binding energies and reduced effective masses that are in excellent agreement with recent reports. Most importantly, suggesting that the combination of the novel dispersion Elliott–Band–Fluctuation model with the Pollmann–B\"{u}ttner model is crucial for accurately determining the reduced effective mass in metal tri-halide perovskites. Hence, we believe this study opens a path for understanding the different aspects governing optical properties in other perovskites systems.
\\
\vskip 7pt
\noindent
\textbf{Acknowledgementements}
\vskip 7pt
\noindent
{K. Lizárraga, P. Venezuela and A. R. Rocha would gratefully like to acknowledge the CAPES, CNPq and FAPESP grants. K. Lizárraga would also like to acknowledge the Pontifical Catholic University of Peru (PUCP) grant no. FAI-065-2023. J. A. Guerra, L. A. Enrique-Morán and E. Serquen acknowledge the support of the Air Force Office of Scientific Research, Grant No. FA9550-25-1-0006 and the Center for Characterization of Materials (CAM-PUCP).}
\vskip 10pt
\noindent
%\textbf{Author contributions}
%\vskip 7pt
%\noindent
%{All authors contribute }
\\
\noindent
\textbf{Conflict of interest}
\vskip 7pt
\noindent
{The authors declare no competing interests.}

\bibliographystyle{aip}
\bibliography{Reference}

\end{document}

\title{Supplementary Information: New Exciton Dispersion Model for the Description the Optical Absorption of 2D Materials}

\author{K. Lizárraga$^{*1,2,3}$}
\author{J. A. Guerra$^{2}$}
\author{L. A. Enrique-Moran$^{2}$}
\author{Cesar E. P. Villegas$^{3}$}
\author{E. Serquen$^{2}$}
\author{E. Ventura$^{2}$}
\author{A. R. Rocha$^4$}
\author{P. Venezuela$^1$}
\affiliation{$^{1}$Instituto de Física, Universidade Federal Fluminense, Rio de Janeiro, Brazil.}
\affiliation{$^{2}$Departamento de Ciencias, Sección Física, Pontificia Universidad Católica del Perú, Lima 32, Peru.}
\affiliation{$^{3}$Departamento de Ciencias, Universidad Privada del Norte, Lima 15434, Peru.}
\affiliation{$^{4}$Instituto de Física Teórica, UNESP, S$\tilde{\textrm{a}}$o Paulo, Brazil.}

\maketitle 
\setcounter{figure}{0}
\renewcommand{\figurename}{Fig.}
\renewcommand{\theequation}{S\arabic{equation}}
\renewcommand{\thefigure}{S\arabic{figure}}
\renewcommand{\thetable}{S\arabic{table}}
\renewcommand{\thesection}{S\arabic{section}}

\section{Fittings and thermal evolution of exciton binding energy, bandgap, exciton broadening and Urbach energy of the optical absorption of metal tri-halide perovskites}

\begin{figure*}[!htb] \centering
\includegraphics[scale=0.38]{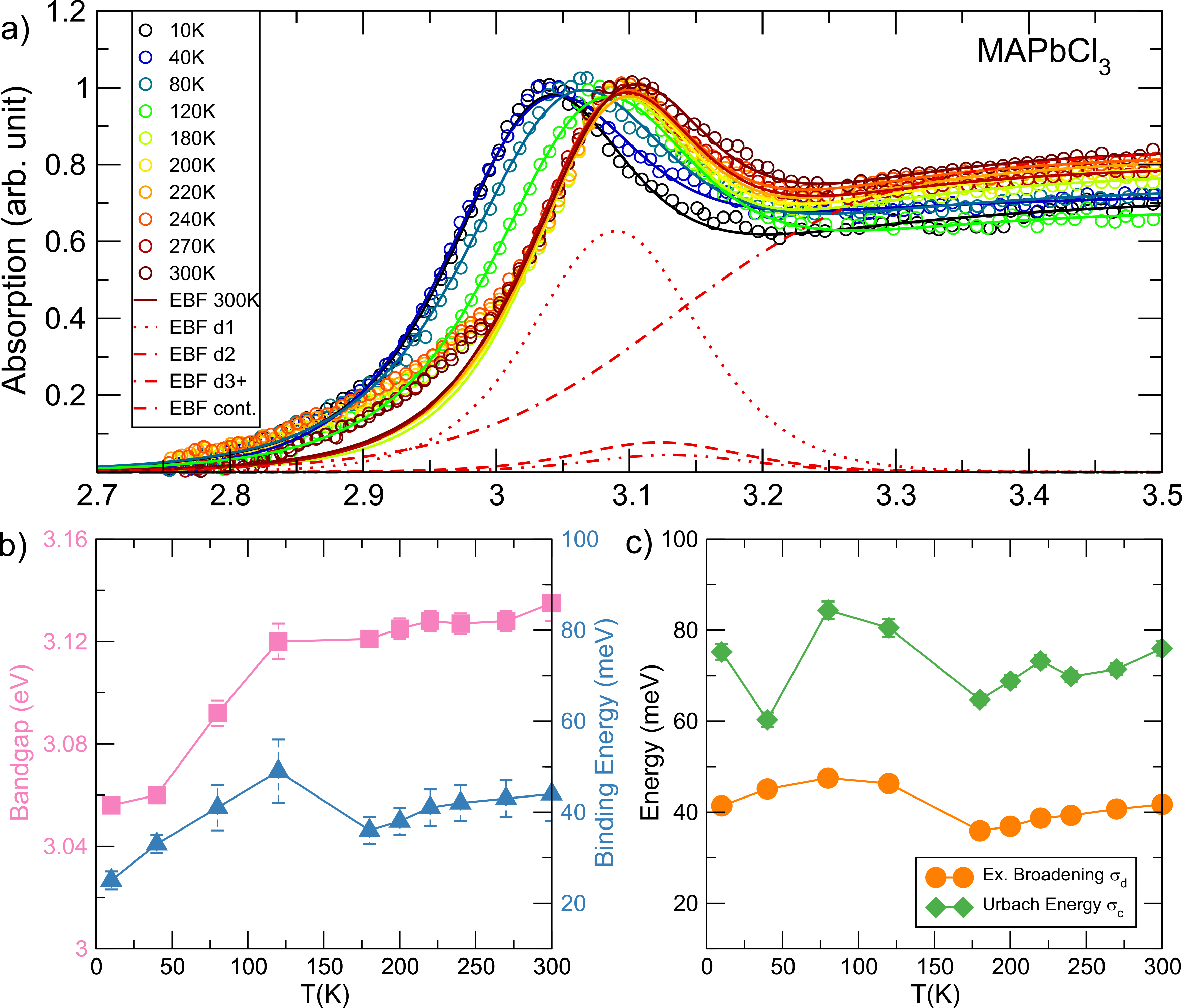}% Here is how to import EPS art
\caption{Fitting of the temperature dependent optical absorption data of MAPbCl$_3$, extracted from Saxena et al. \cite{saxena}, using the EBF model, eq (1) of the main manuscript (a). Thermal evolution of exciton binding energy and bandgap (b). And, thermal evolution of discrete and continuum broadenings of the exciton absorption (c).}
\label{fig:fitebf1} 
\end{figure*}

\begin{figure*}[!htb] \centering
\includegraphics[scale=0.38]{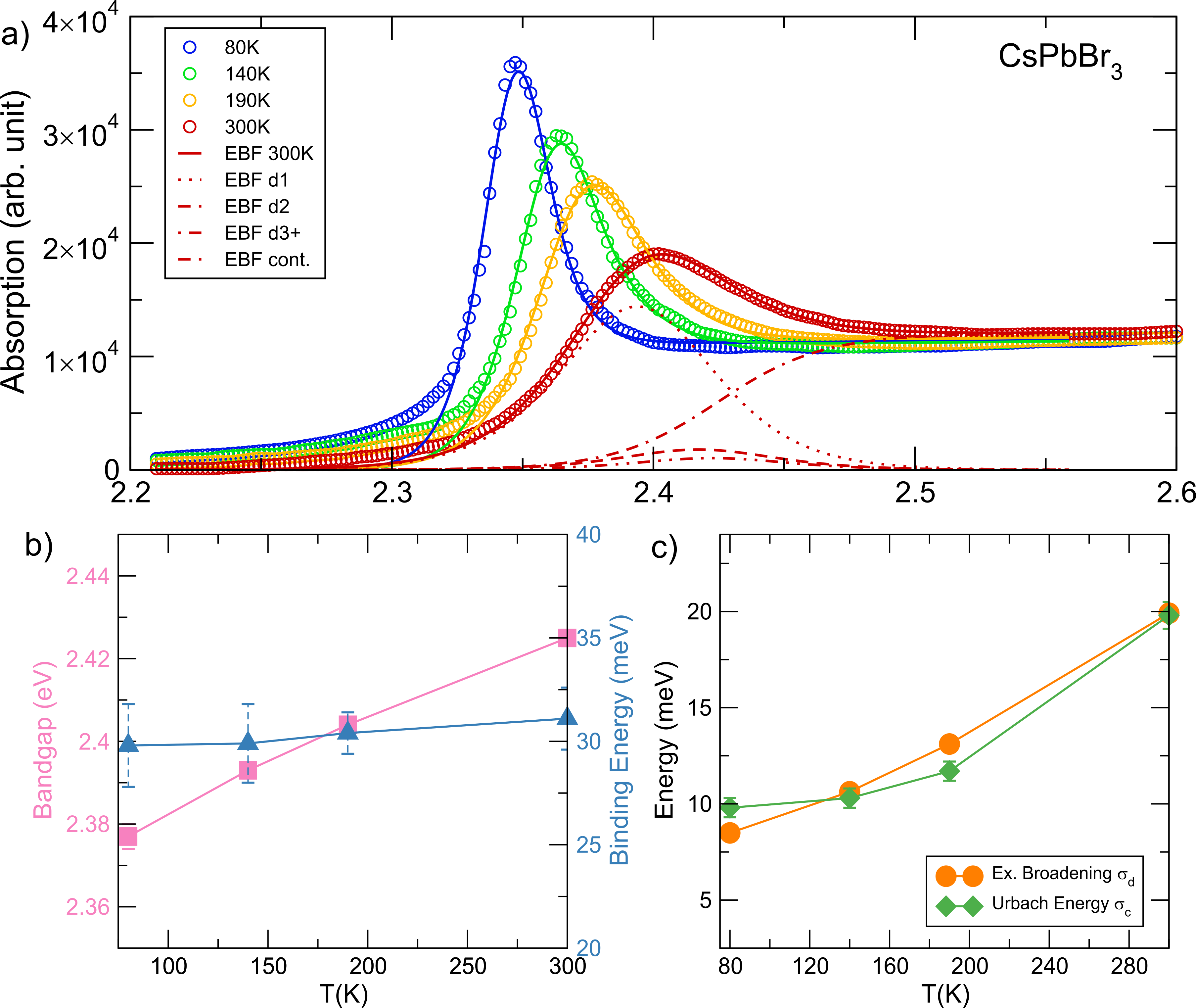}% Here is how to import EPS art
\caption{Fitting of the temperature dependent optical absorption data of CsPbBr$_3$, extracted from Wolf et al. \cite{wolf}, using the EBF model, eq (1) of the main manuscript (a). Thermal evolution of exciton binding energy and bandgap (b). And, thermal evolution of discrete and continuum broadenings of the exciton absorption (c).}
\label{fig:fitebf2} 
\end{figure*}

\begin{figure*}[!htb] \centering
\includegraphics[scale=0.39]{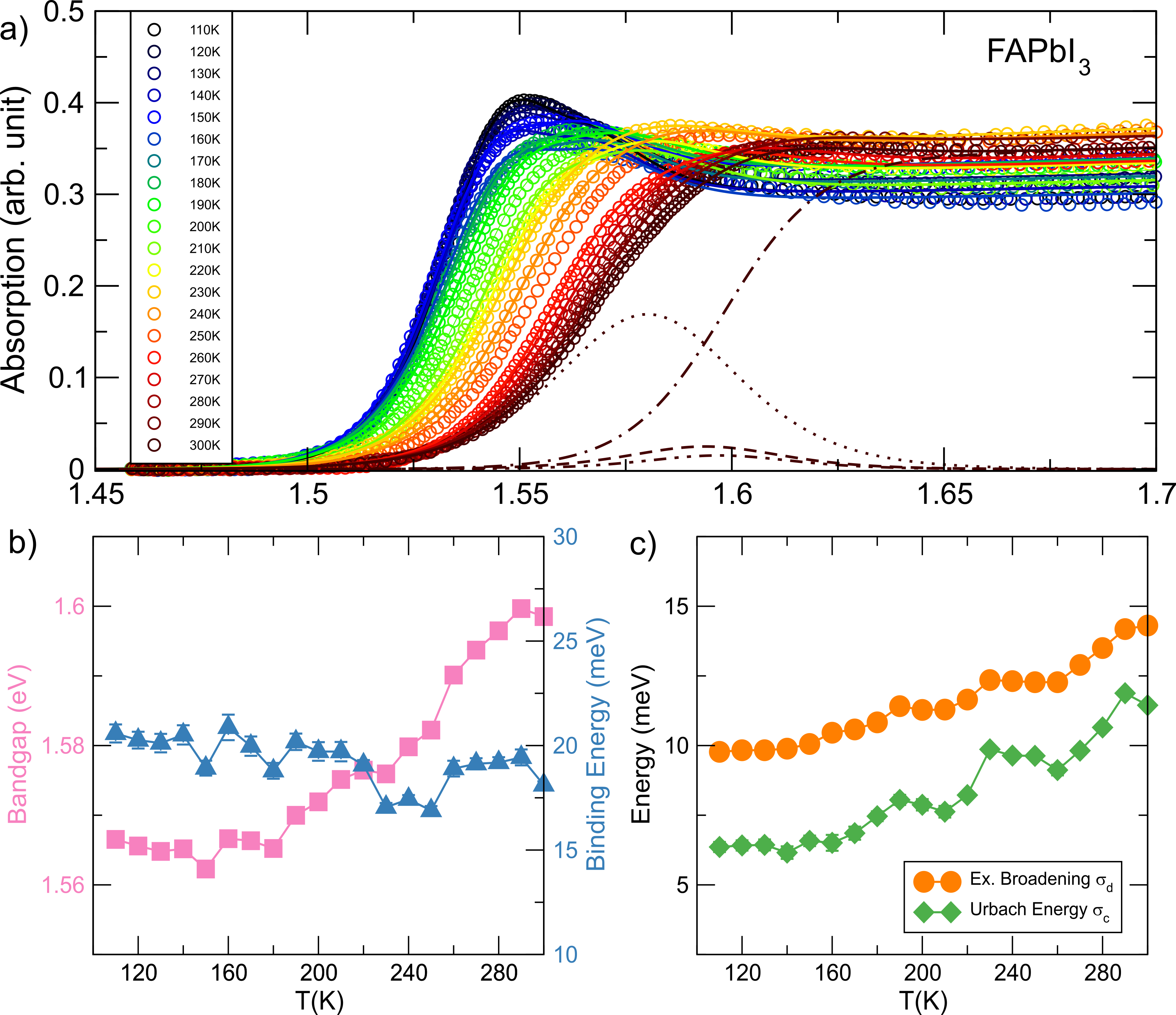}% Here is how to import EPS art
\caption{Fitting of the temperature dependent optical absorption data of FAPbI$_3$, extracted from Xu et al. \cite{youcheng}, using the EBF model, eq (1) of the main manuscript (a). Thermal evolution of exciton binding energy and bandgap (b). And, thermal evolution of discrete and continuum broadenings of the exciton absorption (c).}
\label{fig:fitebf3} 
\end{figure*}

The Best fitted parameters from the EBF model (eq. (1) of the main manuscript) for the optical absorption of MAPbCl$_3$, CsPbBr$_3$ and FAPbI$_3$ are shown in Tables \ref{table:ebf1}, \ref{table:ebf2} and \ref{table:ebf3}, respectively. Fittings of the optical absorption and thermal evolution of the parameters are presented in Figures \ref{fig:fitebf1}, \ref{fig:fitebf2} and \ref{fig:fitebf3}. Absorption data for MAPbCl$_3$, CsPbBr$_3$ and FAPbI$_3$ was extracted from Saxena et al. \cite{saxena}, Wolf et al. \cite{wolf} and Xu et al. \cite{youcheng}, respectively. Note that in the cases of MAPbI$_3$ and MAPbBr$_3$, the best fitted parameters were taken from a previous recent work \cite{lizarraga2}.

\begin{table*}[!htb]
\begin{center}
\begin{tabular}{@{}cccccccc}
\hline\hline
T(K)&A1&A2&$\sigma_{d}$ (meV)&$\sigma_{c}$ (meV)&$E_{g}$ (eV)&$R^{*}$ (meV) \\
\hline
10& 41 (2) & 0.29 (3) & 41.4 (2) & 75.2 (17) & 3.056 (2) & 25 (2)\\
40& 27 (2)& 0.42 (3)& 45.1 (2)& 60.3 (16)& 3060 (2) & 33 (2)\\
80& 22 (4)& 0.49 (7)& 47.5 (2)& 84.4 (18)& 3092 (5)& 41 (5)\\
120& 17 (3)& 0.56 (8)& 46.3 (2)& 80.5 (19)& 3120 (7)& 49 (7)\\
180& 20 (2)& 0.59 (5)& 35.9 (2)& 64.7 (12)& 3121 (3)& 36 (3)\\
200& 19 (3)& 0.63 (6)& 36.9 (2)& 68.8 (13)& 3125 (4)& 38 (3)\\
220& 18 (3)& 0.69 (7)& 38.7 (2)& 73.2 (13)& 3128 (4)&41 (4)\\
240& 18 (2)& 0.69 (7)& 39.3 (2)& 69.8 (12)& 3127 (4)&42 (4)\\
270& 17 (2)& 0.66 (7)& 40.7 (2)& 71.4 (13)& 3128 (4)&43 (4)\\
300& 18 (4)& 0.70 (9)& 41.7 (2)& 76.0 (16)& 3135 (7)&44 (6)\\
  \hline\hline
\end{tabular}
\end{center}
\caption{\label{table:ebf1}$\textrm{MAPbCl}_{3}$ best fitted parameters from the EBF model, eq. (1) of the main manuscript, for temperatures in the range of $10-300$K.}
\end{table*}

%\begin{table*}[!htb]
%\begin{center}
%\begin{tabular}{@{}cccccccc}
%\hline\hline
%T(K)&A1&A2&$\sigma_{d}$ (meV)&$\sigma_{c}$ (meV)&$E_{g}$ (meV)&$R^{*}$ %(meV) \\
%\hline
%10& 40.45 (19) & 0.29 (11) & 41.4 (6) & 75.2 (53) & 3.057 (8) & 25.2 (78) \\
%40& 22.6 (83)& 0.48 (13)& 45.44 (46)& 66.9 (41)& 3069 (11) & 38.8 (100)\\
%80& 21 (17)& 0.52 (32)& 47.77 (54)& 88.5 (63)& 3096 (24)& 43.7 (248)\\
%120& 17 (18)& 0.56 (45)& 46.88 (87)& 87.9 (80)& 3122 (38)& 49.6 (382)\\
%180& 20 (12)& 0.59 (27)& 37.39 (62)& 71.0 (61)& 3123 (16)& 36.7 (157)\\
%200& 19 (14)& 0.63 (35)& 38.58 (59)& 76.0 (66)& 3126 (20)&38.5 (198)\\
%220& 17 (12)& 0.68 (34)& 39.48 (74)& 76.6 (56)& 3129 (20)&41.9 (195)\\
%240& 17 (10)& 0.68 (28)& 40.05 (73)& 72.6 (51)& 3127 (17)&42.1 (169)\\
%270& 17 (9)& 0.66 (24)& 40.71 (64)& 71.4 (46)& 3128 (16)&42.8 (149)\\
%300& 17 (12)& 0.70 (35)& 41.73 (81)& 76.6 (56)& 3135 (22)&43.6 (213)\\
%  \hline\hline
%\end{tabular}
%\end{center}
%\caption{\label{table:parameters}$\textrm{MAPbCl}_{3}$ best fitted parameters of eq. () of the main manuscript for temperatures in the range of 10-300K}
%\end{table*}

\begin{table*}[!htb]
\begin{center}
\begin{tabular}{@{}cccccccc}
\hline\hline
T(K)&A1&A2&$\sigma_{d}$ (meV)&$\sigma_{c}$ (meV)&$E_{g}$ (eV)&$R^{*}$ (meV) \\
\hline
80& 260269 (25684) & 0.58 (3) & 8.49 (10) & 9.8 (5) & 2.377 (3) & 29.8 (20)\\
140& 260269 (20066)  & 0.60 (3) & 10.63 (13) & 10.3 (6) & 2.393 (2) & 29.9 (19)\\
190& 260269 (10443) & 0.61 (2) & 13.11 (11) & 11.7 (5) & 2.404 (1) & 30.4 (10)\\
300& 250269 (14610) & 0.67 (3) & 19.91 (13) & 19.8 (7) & 2.425 (2) & 31.1 (15)\\
  \hline\hline
\end{tabular}
\end{center}
\caption{\label{table:ebf2}$\textrm{CsPbBr}_{3}$ best fitted parameters from the EBF model, eq. (1) of the main manuscript, for temperatures in the range of $80-300$K}
\end{table*}

\begin{table*}[!htb]
\begin{center}
\begin{tabular}{@{}cccccccc}
\hline\hline
T(K)&A1&A2&$\sigma_{d}$ (meV)&$\sigma_{c}$ (meV)&$E_{g}$ (eV)&$R^{*}$ (meV) \\
\hline
110& 3.59 (7) & 0.93 (1) & 9.77 (7) & 6.36 (18) & 1.567 (1) & 20.58 (43)\\
120& 3.60 (6) & 0.98 (1) & 9.83 (9) & 6.42 (18) & 1.566 (1) & 20.26 (40)\\
130& 3.55 (7) & 0.99 (1) & 9.83 (9) & 6.44 (20) & 1.565 (1) & 20.12 (44)\\
140& 3.33 (6) & 1.01 (1) & 9.89 (10) & 6.16 (22) & 1.565 (1) & 20.51 (46)\\
150& 3.59 (6) & 1.05 (1) & 10.05 (10) & 6.57 (19) & 1.562 (1) & 18.92 (36)\\
160& 3.17 (8) & 1.03 (1) & 10.46 (14) & 6.51 (29) & 1.566 (1) & 20.86 (61)\\
170& 3.29 (6) & 1.07 (1) & 10.57 (12) & 6.86 (24) & 1.566 (1) & 19.97 (46)\\
180& 3.52 (6) & 1.08 (1) & 10.83 (10) & 7.46 (18) & 1.565 (1) & 18.79 (38)\\
190& 3.35 (5) & 1.10 (1) & 11.41 (10) & 8.05 (18) & 1.570 (1) & 20.18 (39)\\
200& 3.29 (6) & 1.10 (1) & 11.28 (11) & 7.87 (20) & 1.572 (1) & 19.72 (44)\\
210& 3.22 (6) & 1.09 (1) & 11.29 (11) & 7.63 (21) & 1.575 (1) & 19.71 (46)\\
220& 3.34 (3) & 1.13 (1) & 11.65 (6) & 8.22 (12) & 1.576 (0) & 19.71 (23)\\
230& 3.72 (3) & 1.18 (1) & 12.35 (5) & 9.86 (10) & 1.576 (0) & 17.05 (20)\\
240& 3.50 (3) & 1.23 (1) & 12.31 (6) & 9.64 (10) & 1.580 (0) & 17.43 (20)\\
250& 3.45 (3) & 1.26 (1) & 12.27 (6) & 9.63 (10) & 1.582 (0) & 16.89 (22)\\
260& 3.10 (4) & 1.25 (1) & 12.27 (11) & 9.12 (19) & 1.590 (1) & 18.90 (38)\\
270& 3.10 (3) & 1.27 (1) & 12.89 (8) & 9.82 (14) & 1.594 (1) & 19.13 (27)\\
280& 3.10 (3) & 1.30 (1) & 13.50 (7) & 10.65 (11) & 1.596 (1) & 19.19 (26)\\
290& 3.10 (4) & 1.34 (1) & 12.18 (8) & 11.87 (14) & 1.599 (1) & 19.43 (38)\\
300& 3.14 (3) & 1.32 (1) & 14.31 (7) & 11.45 (11) & 1.599 (0) & 18.11 (18)\\
  \hline\hline
\end{tabular}
\end{center}
\caption{\label{table:ebf3}$\textrm{FAPbI}_{3}$ best fitted parameters from the EBF model, eq. (1) of the main manuscript, for temperatures in the range of $110$-$300$K in steps of 10K}
\end{table*}

\clearpage
\section{Determination of $\varepsilon_{\infty}$ through Cauchy's model and previous reports}

\begin{figure*}[!htb] \centering
\includegraphics[scale=0.50]{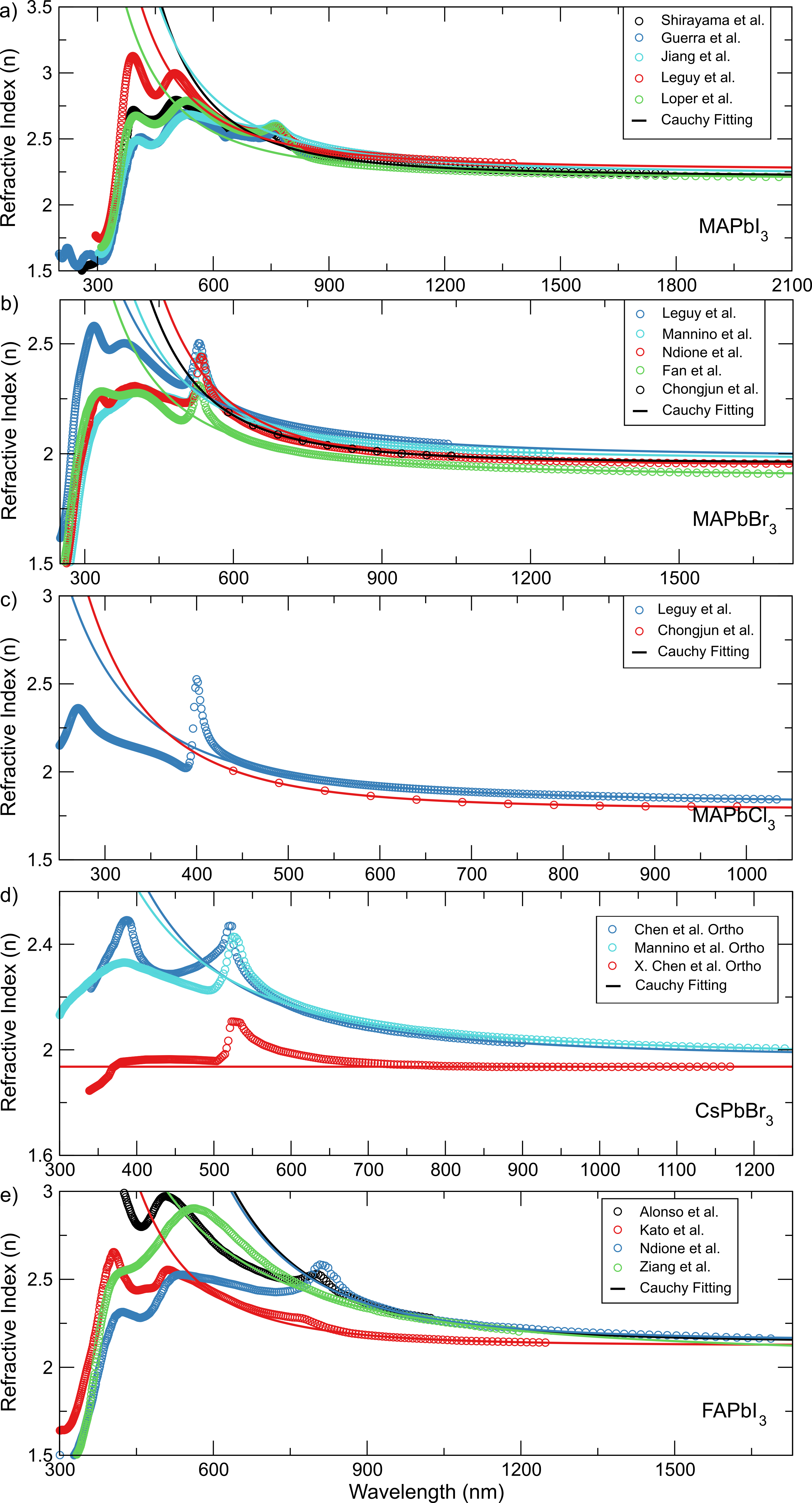}% Here is how to import EPS art
\caption{Fitting of the refractive index using the Cauchy's model, eq. (9) of the main manuscript, for MAPbI$_3$ (a), MAPbBr$_3$ (b), MAPbCl$_3$ (c), CsPbBr$3$ (d) and FAPbI$_3$ 
(e). Note that the data collected is at 300K where MAPbI$_3$ is tetragonal, CsPbBr$_3$ is orthorhombic, and MAPbBr$_3$, MAPbCl$_3$ and FAPbI$_3$ are cubic.}
\label{fig:einfmaperovs} 
\end{figure*}

Eq. (5) of the main manuscript was used to fit the refractive index, $n$, in the transparent region from previous literature reports. Table \ref{table:cauchy} show the references from which data were extracted along with the best fitted parameters of eq. (5). Figure \ref{fig:einfmaperovs} shows the fittings of the refractive index for the perovskite systems. Note that the phases of perovskites was selected at 300K. 

Experimental (E) and theoretical (T) literature reports of $\varepsilon_{\infty}$ are presented in Table \ref{table:einf1} along with their corresponding methodologies and models used. Note that the phase of the perovskite is denoted in the superscripts: $^O$, $^T$ and $^C$ highlight the orthorhombic, tetragonal and cubic phases, respectively. 

\begin{table*}[ht]
\begin{center}
\begin{tabular}{@{}m{2.5cm} m{2.0cm} m{2.0cm} m{2.0cm} m{2.0cm} m{3.0cm}}
\hline\hline
Material & $n_{\infty}$ & $N_1$ & $N_{2}$ & $\varepsilon_\infty=n_{\infty}^2$ & Extracted Data Ref.\\
\hline
MAPbI$_3$ 
& 2.28 (0)   & 272 (7) & 7526 (30) & 5.2 (0) & Alonso et al. \cite{isabel} \\ 
& 2.25 (0)   & - & - & 5.1 (0)  
& Guerra et al. \cite{guerra3} \\ 
(Tetragonal)  & 2.26 (0)   & 903 (6) & 2104 (33) & 5.1 (0)  
& Leguy et al. \cite{leguy2} \\
& 2.23 (0)   & 1283 (6) & 3051 (22) & 5.0 (0)  
& Jiang et al. \cite{jiang} \\ 
& 2.20 (0)   & 974 (2) & 1209 (4) & 4.8 (0)  
& Loper et al. \cite{loper} \\
& 2.21 (0)   & 660 (1) & 4695 (3) & 4.9 (0)  
& Shirayama et al. \cite{shirayama} \\ 
\hline 
MAPbBr$_3$ & 1.98 (0)   & 659 (4) & 542 (8) & 3.9 (0) & Leguy et al.\cite{leguy2}\\
(Cubic) & 1.97 (0)   & 409 (4) & 1219 (20) & 3.9 (0) & Mannino et al. \cite{mannino}\\
& 1.96 (0)   & 129 (26) & 2360 (42) & 3.8 (0) & Chongjun et al. \cite{chongjun}\\
& 1.95 (0)   & 156 (2) & 3053 (4) & 3.8 (0) & Ndione et al. \cite{ndione}\\
& 1.89 (0)   & 488 (4) & 698 (5) & 3.6 (0) & Fan et al. \cite{yubifan}\\
\hline
MAPbCl$_3$ & 1.82 (0)   & 258 (2) & 394 (2) & 3.3 (0) & Leguy et al. \cite{leguy2} \\
(Cubic)& 1.78 (0)   & 58 (12) & 719 (13) & 3.2 (0) & Chongjun et al. \cite{chongjun} \\
\hline
CsPbBr$_3$ & 1.97 (0)   & 310 (7) & 1349 (12) & 3.9 (0) & Chen et al. \cite{chenchen}\\ 
(Ortho.)& 1.97 (0)   & 474 (4) & 890 (21) & 3.9 (0) & Mannino et al. \cite{mannino}\\ 
& 1.94 (0)   & - & - & 3.8 (0) & X. Chen et al. \cite{xiaoxuan}\\ 
%& \crule{5} \\
%(Cubic) & 1.70 (0)   & 369 (5) & 1042 (6) & 2.9 (0) & Washeng et al. \cite{washeng}\\
% & 1.63 (0)   & 583 (3) & - & 2.7 (0) & Zhao et al. \cite{minglinzhao}\\ 
\hline
FAPbI$_3$ & 2.14 (0)   & - & 14477 (7) & 4.6 (0) & Alonso et al. \cite{isabel}\\
(Cubic) & 2.15 (0)   & - & 13772 (5) & 4.6 (0) & Ndione et al. \cite{ndione}\\ 
& 2.12 (0)   & - & 3844 (6) & 4.5 (0) & Kato et al. \cite{fujiwara}\\ 
& 2.05 (0)   & 2243 (7) & 605 (30) & 4.2 (0) & Ziang et al. \cite{ziang}\\ 
  \hline\hline
\end{tabular}
\end{center}
\caption{\label{table:cauchy} Best fitted parameters of the Cauchy model, eq. (9) of the main manuscript, to $\varepsilon_1$ in the transparent region of literature reports.}
\end{table*}

\begin{table}[!htb]
\begin{center}
\begin{tabular}{@{}m{2.5cm} m{1.5cm} m{5.5cm} m{4.5cm} m{1.0cm}}
\hline\hline
Material & $\epsilon_{\infty}$ & Methodology & Model & Ref. \\
 \hline
 $\textrm{MAPbI}_{3}$  &  6.5$^{O}$ & Diffuse Reflectivity (4.2K)  & - &\cite{hirasawa}  \\
 &  5.0$^{T}$ & Far IR Spectroscopy (300K) & TL osc. &\cite{glaser}\\
 &  5.5$^{T}$ & THz Spectroscopy (300K) & TL $\&$ Drude osc. &\cite{valverdechavez}\\
                      & 5.6$^{T}$& Ellipsometry at 1.21eV (300K) & TL $\&$ Crit. Point &\cite{isabel} \\
                      & 5.5$^{T}$ & Ellipsometry at 1eV (300K) & Critical Point &\cite{leguy2} \\ 
                      E & 5.1$^{T}$  & Far IR Spectroscopy (300K) & TL $\&$ Drude osc.& \cite{soufiani} \\         & 5.1$^{T}$& Ellipsometry (300K) & Point-by-point $\&$ Cauchy &\cite{guerra3} \\ 
                      & 5.1$^{T}$ & Ellipsometry (300K) & Cauchy &\cite{jiang} \\
                    & 4.9$^{T}$ & Ellipsometry (300K) & TL osc. &\cite{shirayama} \\
                      & 4.0$^{T}$ & Ellipsometry (300K) & Forouhi-Bloomer &\cite{loper} \\ 
%                      &  4.5$^{C}$ & QSGW with SOC (C) &\cite{brivio} \\
%                      & 5.8$^{C}$  & HSE with SOC (C) &\cite{bah} \\
                       \cline{2-5} 
%& 6.8$^{C}$ & DFPT & GW with SOC &\cite{bokdam} \\
                      & 5.6$^{T}$ & DFT & HSE with SOC &\cite{bah} \\
%                      & 6.6$^{T}$& SR-DFT (T)&\cite{umari2} \\
                      & 5.6$^{T}$ & DFPT & GW with SOC &\cite{umari} \\
                     T & 5.3$^{T}$ & DFT with $\alpha=1/\varepsilon_\infty$ & PBE0$\alpha$ with SOC &\cite{menendez2} \\
                     & 5.1$^{T}$ & DFT & HSE with SOC &\cite{du} \\
                     & 5.0$^{T}$ & DFT & HSE &\cite{bah} \\
%                      & 6.23$^{T}$ & DFPT (T) &\cite{leveillee} \\
%                      & 6.24$^{C}$ & DFPT (C) &\cite{leveillee} \\
%                      & 6.83$^{C}$ & DFPT (C) &\cite{bokdam} \\
%                      & 7.34$^{T}$ & DFPT w/wo SOC &\cite{saxena} \\
                      \hline\hline
$\textrm{MAPbBr}_{3}$ &  4.8$^{O}$ & Absolute Reflectivity (4.2K) & - &\cite{tanaka}  \\
                       & 4.7$^{C}$ & Far IR Spectroscopy (300K) & TL osc. &\cite{glaser}\\
                       & 4.2$^{C}$ & Ellipsometry at 1.2eV (300K) & Critical Point &\cite{leguy2} \\
                      E &4.0$^{C}$ & Ellipsometry at 1eV (300K) & Critical Point &\cite{mannino} \\
                      & 4.0$^{C}$ & Transmittance at 1.2eV (300K) & Brewster's law &\cite{chongjun} \\
                      & 3.9$^{C}$ & Ellipsometry at 0.72eV (300K) & TL osc. &\cite{ndione} \\
                      & 3.6$^{C}$ & Ellipsometry at 0.73eV (300K) & - &\cite{yubifan} \\  
                      \cline{2-5} 
                      %& 5.2$^{C}$ & DFPT & GW with SOC &\cite{bokdam} \\
                       & 4.4$^{C}$  & DFT & HSE with SOC&\cite{bah}\\
                      T & 4.2$^{C}$ & DFT+U & PBE with SOC &\cite{welch} \\
                      & 4.1$^{C}$ & DFT & HSE &\cite{bah} \\
                      & 3.8$^{C}$ & DFT & PBE with vdW &\cite{ungi} \\
\hline\hline
$\textrm{MAPbCl}_{3}$ & 4.0$^C$ & Far IR Spectroscopy (300K) & TL osc. &\cite{glaser} \\
                      E & 3.4$^C$ &  Ellipsometry at 1.2eV (300K) & Critical Point &\cite{leguy2} \\
                      & 3.3$^C$ & Transmittance at 1.3eV (300K) & Brewster's law &\cite{chongjun} \\
                      \cline{2-5} 
                      %& 4.2$^C$ & DFPT & GW with SOC &\cite{bokdam} \\
%                      &  3.8$^C$  & DFT & HSE with SOC &\cite{bah}\\
                      & 3.5$^{C}$ & DFT & HSE &\cite{bah} \\
                      T&  3.4$^C$ & DFT+U & PBE with SOC &\cite{welch} \\
                      & 3.1$^{C}$ & DFT & PBE with vdW &\cite{ungi2} \\
\hline\hline
$\textrm{CsPbBr}_{3}$ & 4.0$^O$ & Ellipsometry at 1.38eV (300K) & TL $\&$ Cauchy & \cite{chenchen} \\
                    E  & 4.0$^{O}$  & Ellipsometry at 1eV (300K)  & Critical Point &\cite{mannino} \\
                     &3.7$^O$ & Ellipsometry at 1eV (300K) & TL osc. &\cite{xiaoxuan} \\
%                      &3.1$^C$ & Ellipsometry at 1.13eV & TL $\&$ Gauss osc. &\cite{washeng} \\
%                    &3.0$^C$ & Ellipsometry at 1.58eV & TL osc. &\cite{minglinzhao} \\
%                      &  4.7$^C$  & DFPT with SOC (C) &\cite{wolf}\\
                     \cline{2-5} 
                     %& 4.5$^O$ & DFPT & GW with SOC &\cite{filip} \\
                      %& 4.3$^O$ & DFT & EV-GGA with SOC &\cite{ghaithan} \\
                      & 3.9$^O$ & DFT & mBJ-GGA with SOC &\cite{ghaithan} \\
                      T & 3.8$^O$ & DFT & PBE & \cite{sapori} \\
                      & 3.8$^O$ & DFT & LDA & \cite{tomanova} \\
\hline\hline
$\textrm{FAPbI}_{3}$ & 5.2$^C$ & Ellipsometry at 1.2eV (300K) & TL $\&$ Crit. Point &\cite{isabel} \\
                & 5.1$^C$ & Ellipsometry at 0.73eV (300K) & Critical Point & \cite{subedi} \\
                & 4.8$^C$ & Ellipsometry at 1eV (300K) & Forouhi-Bloomer & \cite{ziang} \\
               E & 4.7$^C$ & Ellipsometry at 0.72eV (300K) & TL osc. & \cite{ndione} \\
                & 4.6$^C$ & Ellipsometry at 1eV (300K) & TL osc. & \cite{fujiwara} \\                     
%                      & 5.0$^H$ & Ellipsometry at 1eV (300K) & TL osc. & \cite{fujiwara} \\
%                      & 3.7$^H$ & Ellipsometry at 1.2eV (300K) & TL $\&$ Crit. Point & \cite{isabel} \\ 
                      \cline{2-5} &  5.5$^C$ & DFT & mBJ-GGA &\cite{ghtami}\\ 
                      T &  5.3$^C$ & DFPT & GW with SOC &\cite{zeeshan}\\
                       &  5.2$^C$ & DFT & PBE &\cite{Mayengbam}\\
                      &  4.6$^T$ & DFPT & GW with SOC &\cite{zeeshan}\\
%                      &  4.0$^H$ & DFPT & GW with SOC &\cite{zeeshan}\\
%                    \hline
%$\textrm{MAPbI}_{2}$Cl$_1$ &  5.07 & &\cite{welch}\\
  \hline\hline
\end{tabular}
\end{center}
\caption{Experimental (E) and theoretical (T) literature reports of the electronic (optical or high frequency) dielectric constant of MAPbI$_3$, MAPbBr$_3$, MAPbCl$_3$, CsPbBr$_3$ and FAPbI$_3$. Methodologies and model used are shown as well. Units are in terms of the vacuum permittivity. Note that the phase of perovskite are also reported as superscripts $^O$, $^T$ and $^C$ for the orthorhombic, tetragonal and cubic phases, respectively.\label{table:einf1}}
\end{table}

\clearpage

\section{Determination of $\hbar\omega_{LO}$ through the modeling of the Exciton Broadening.}

\begin{figure*}[!htb] \centering
\includegraphics[scale=0.45]{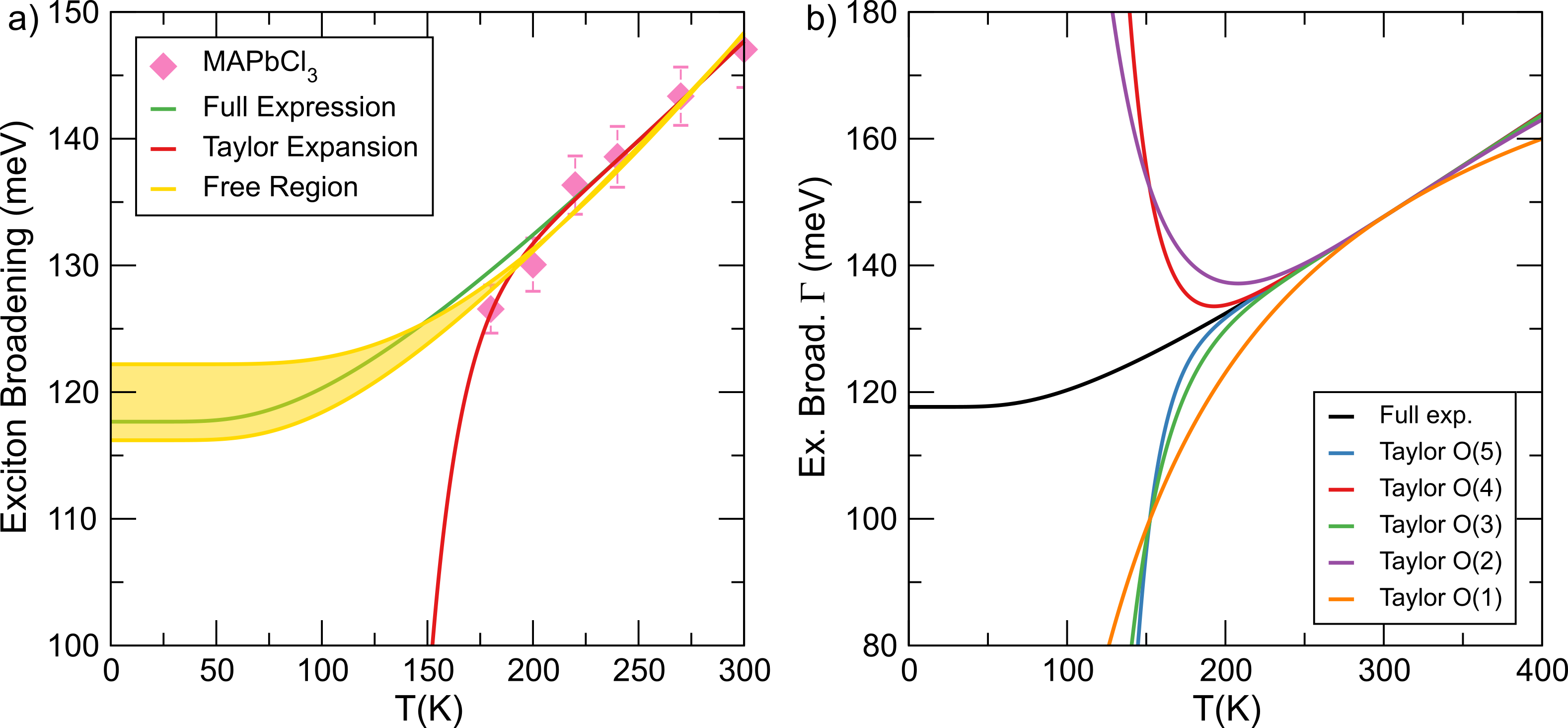}% Here is how to import EPS art
\caption{(a) Fitting of the exciton broadening of MAPbCl$_3$ using eq. (7) of the main manuscript (Full expression) and eq. (\ref{eq.gamma}) from the Taylor expansion. (b) Comparison between the full expression, eq. (7), the Taylor expansion terms of eq. (\ref{eq.gamma}).}
\label{fig:hwlo1} 
\end{figure*}

The fitting procedure of eq. (7) is detailed in the following. When eq. (7) is applied to model the exciton broadening of the perovskites at a given phase (e.g., cubic MAPbCl$_3$ that ranges temperatures of $180-300$K), the absence of data points at low temperatures $-$caused by the discontinuity of the phase transition$-$, produce a large bias on the parameter determination of eq. (7). For instance, the shaded region in Figure \ref{fig:hwlo1}a shows all the accessible trends for MAPbCl$_3$. To solve this problem, we have expanded eq. (7) using the Taylor series up to the $5^{th}$ order, i.e.,
\begin{center}
\begin{align}
    \Gamma(T)_{x\rightarrow1}=&\Gamma_0 + \Gamma_{LO} \Bigg(  \frac{1}{e-1} - \frac{e(x-1)}{(e-1)^2} +   \frac{e(e+1)(x-1)^2}{2(e-1)^3} -\frac{e(e^2+4e+1)(x-1)^3}{6(e-1)^4}  \nonumber \\ 
     &+ \frac{e(e^3+11e^2+11e+1)(x-1)^4}{24(e-1)^5} - \frac{e(e^4+26e^3+66e^2+26e+1)(x-1)^5}{120(e-1)^6} + O(6) \Bigg), \nonumber \\
    x=&\frac{\hbar\omega_{LO}}{k_\beta T}.\label{eq.gamma}
\end{align}
\end{center}

\begin{figure*}[!htb] \centering
\includegraphics[scale=0.45]{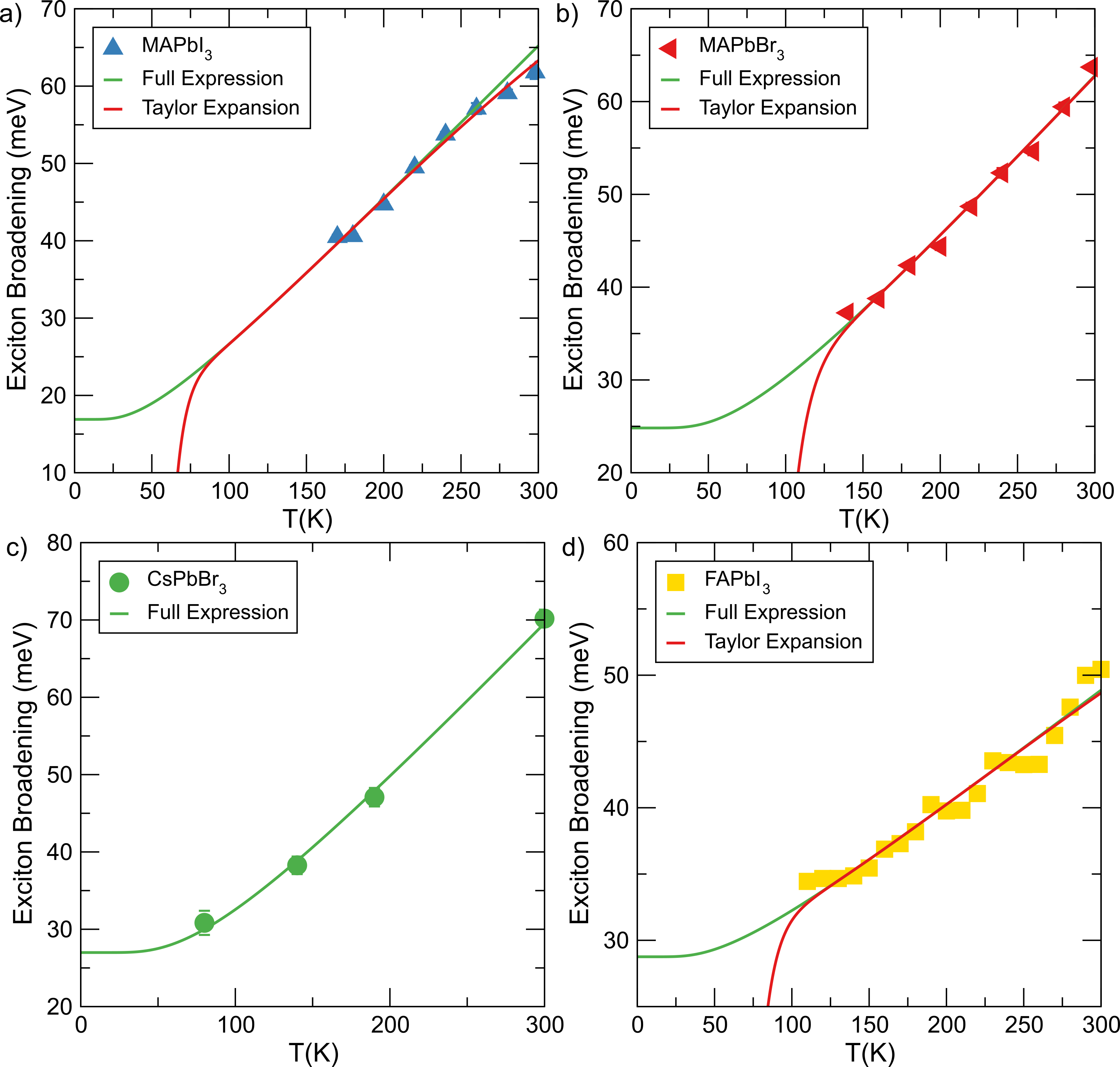}% Here is how to import EPS art
\caption{Fitting of the exciton broadening of MAPbI$_3$ (a), MAPbBr$_3$ (b), CsPbBr$_3$ (c), FAPbI$_3$ (d), using eq. (7) of the main manuscript (Full expression) and eq. (\ref{eq.gamma}) from the Taylor expansion.}
\label{fig:hwlo2} 
\end{figure*}

Eq. (\ref{eq.gamma}) is then used to fit the thermal evolution of the exciton broadening for the halide perovskites. A comparison of the full expression of eq. (7) along with the different expansion terms of eq. (\ref{eq.gamma}) is presented in Figure \ref{fig:hwlo1}b. And, the fitting results for MAPbCl$_3$ and the rest of halide perovskite systems are presented in Figures \ref{fig:hwlo1}a and \ref{fig:hwlo2}, respectively. The best fitted parameters of eq. (\ref{eq.gamma}) are summarized in Table \ref{table:linewidth}.

\begin{table*}[ht]
\begin{center}
\begin{tabular}{@{}lcccc}
\hline\hline
Material & $\Gamma_{0}$(meV) & $\Gamma_{LO}$(meV) & $\hbar\omega_{LO}$(meV) &  Raman Highest LO Mode (meV) \\
\hline
MAPbI$_3$ & 16 (9)   & 24 (5) & 10.0 (41) & 17.5$^T$\cite{leguy} \\
MAPbBr$_3$ & 25 (4)   & 38 (12) & 17.8 (51) & 18.6$^{T/C}$\cite{leguy}\\
MAPbCl$_3$ & 118 (3)   & 53 (4) & 26.2 (12) & 29.5$^C$\cite{leguy}\\
CsPbBr$_3$ & 27 (3)   & 48 (27) & 19.5 (86) & 18.5$^O$\cite{keisei}, 20.4$^O$\cite{stoumpos} \\
FAPbI$_3$ & 29 (3)   & 15 (6) & 14.4 (60) &  14.1$^C$\cite{josefa}, 16.9$^C$ \cite{steele} \\
  \hline\hline
\end{tabular}
\end{center}
\caption{\label{table:linewidth} Best fitted parameters of the thermal evolution of the exciton broadening from eq. (\ref{eq.gamma}) for the halide perovskite systems. Raman cage modes corresponding to the highest LO phonon energy are shown as well for comparison purposes only. Superscripts denotes the perovskites phases: orthorhombic (O), tetragonal (T) and cubic (C).}
\end{table*}

\clearpage

\section{Reports of Carrier Bare Effective Masses}
\begin{table*}[!htb]
\centering
\begin{tabular}{@{}m{2.0cm} m{1.5cm} m{1.5cm} m{1.5cm} m{6.5cm}  m{1.0cm}}
\hline\hline
Material & $m^{*}_e$ & $m^{*}_h$ & $\mu$ & Methodology  & Ref. \\
 \hline
$\textrm{MAPbI}_{3}$   &   &  & 0.150$^O$ & Magnetoabsorption (5K)  &       \cite{tanaka}  \\
                      &   &  & 0.120$^O$ & Magnetoabsorption (4.2K) &    \cite{hirasawa}  \\
                        &   &  & 0.109$^O$ & High Field Magnetoabsorption &  
                      \cite{soufiani3}  \\
                    &    &   &  0.104$^O$ & High Field Magnetoabsorption  &\cite{galkowski} \\
%                    &   &  & 0.300$^T$ & Transient Abs.(RT) $\&$ Burstein-Moss Eff.   &  \cite{manser}  \\
                   E &   &  & 0.140$^T$ & Transient Abs.(RT) $\&$ Burstein-Moss Eff.   &  \cite{price}  \\
                      &   &  & 0.109$^T$ & High Field Magnetoabsorption & 
                      \cite{soufiani3}  \\
                    &    &   &  0.104$^T$ & High Field Magnetoabsorption  &\cite{galkowski} \\
                      &    &   &  0.104$^T$ & High Field Magnetoabsorption &  \cite{miyata} \\
                      &    &   &  0.104$^T$ & T and $\rho$-resolved Optical Spect. (RT)&  \cite{yangliu} \\
                    &    &   &  0.100$^T$ & Microscopic Optical Detection &  \cite{andrew} \\
 %                   &    &  0.50(10) & $^{C/T}$ & LEED-ARPES (RT) (along X-R)  & \cite{fengzhou}\\
                    &    &  0.35(15) & $^{C/T}$  & LEED-ARPES (RT) (along $\Gamma$-X)  & \cite{jinpeng}\\
                    &    &  0.24(10) & $^{C/T}$  & LEED-ARPES (RT) (along $\Gamma$-M)  & \cite{jinpeng}\\
                    &    &  0.18(6) & $^{C/T}$  & LEED-ARPES (RT)  (along X-M)  & \cite{fengzhou}\\
%                    &  0.11 & 0.13 & 0.060$^O$ & DFT with SOC    &  \cite{marina} \\
%                     \cline{2-6} &  0.16 & 0.18 & 0.085$^O$ & DFT with GW and SOC  &     \cite{marina} \\
%                      &  0.22 & 0.23 & 0.112$^O$ & DFT with GW  &     \cite{marina} \\
%                        &  0.19 & 0.23 & 0.104$^O$ & DFT with SOC   &  \cite{menendez}\\
                      \cline{2-6} &  0.18  & 0.24 & 0.103$^T$ & DFT with HSE and SOC   &   \cite{bah} \\
                      &  0.23 & 0.33 & 0.117$^T$ & DFT with SOC   &    \cite{mosconi} \\
                     T &  0.17 & 0.28 & 0.106$^T$ & DFT with SOC (Avg. values)    &  \cite{umari} \\
%                      &  0.26 & 0.44 & 0.163$^T$ & DFT %with SOC  $\Gamma$-Z  &   \cite{umari} \\
%                      &  0.14 & 0.18 & 0.079$^T$ & DFT with HSE and SOC  &    \cite{yingbo} \\
%                      &  0.19 & 0.25 & 0.108$^T$& DFT  with GW and SOC (Avg. values)   &    \cite{umari} \\
%                      &  0.29 & 0.40 & 0.168$^T$& DFT with GW and SOC  $\Gamma$-Z  &    \cite{umari} \\
%                     &  0.28 & 0.26 & 0.134$^T$ & Scalar Relativistic DFT  $\Gamma$-R  &   \cite{umari} \\
%                     &  0.23 & 0.29 & 0.128$^T$ & DFT ($\Gamma$-Z)   &  \cite{Berdiyorov} \\
%                      T &  0.37 & 0.28 & 0.159$^T$ & DFT with SOC  &    \cite{ali} \\
                      &  0.17  &  0.23 &  0.098$^C$  & DFT with GW and SOC &  \cite{bokdam} \\ 
%                      &  0.15  & 0.12 & 0.067$^C$ & DFT with HSE and SOC  &   \cite{brivio} \\
%                      &  0.13  & 0.17 & 0.074$^C$ & DFT with HSE and SOC  &    \cite{bah} \\                    
%                      &  0.32 & 0.36 & 0.169$^C$ & DFT  &  \cite{giorgi} \\
%                      &  0.23 & 0.29 & 0.128$^C$ & DFT with SOC  &    \cite{giorgi} \\
%                      &  0.23 & 0.24 & 0.117$^C$ & DFT with SOC   &   \cite{jacky} \\                    
%                      &  0.19 & - & & DFT with SOC   &    O &  \cite{leveillee} \\
%                      &  0.16 & - & & DFT with SOC  &  T &  \cite{leveillee} \\
%                      &  0.23 & - & & DFT with SOC  &  C &  \cite{leveillee} \\
%                      &  0.73 & 0.36 & & Scalar Relativistic DFT Avg.  &  T &  \cite{umari} \\
%                      &  0.40 & 0.36 & & Scalar Relativistic DFT  $\Gamma$-A  &  T &  \cite{umari} \\
%                      &  0.23 & 0.25 & 0.120$^C$ & DFT with SOC  &    \cite{choljun} \\
%                      &  0.18 & 0.19 & 0.092$^C$ & DFT (Virtual Crystal Approx.)  &   \cite{ungi} \\
%                      &  0.19 & 0.24 & 0.106$^C$ & DFT &    \cite{ungi2} \\
%                      &  0.33 & 0.37 & & DFT  &  C &  \cite{nael} \\
                      &  0.20 & 0.29 & 0.118$^C$ & DFT with SOC   &     \cite{nael} \\
%                      &  0.38 & 0.51 & & DFT with U  &   C &  \cite{welch} \\
%                      &  0.26 & 0.46 & & DFT with U (Less.e-)  &   C &  \cite{welch} \\
%                      &  0.30 & 0.25 & 0.136$^C$ & DFT  &    \cite{wangk} \\
%                      &  0.20 & 0.23 & 0.107$^C$ & DFT with Van Der Waals  &  \cite{geng} \\
%                      &  0.18 & 0.26 & 0.106$^C$ & DFT  &     \cite{saxena} \\
                      \hline\hline
$\textrm{MAPbBr}_{3}$  &   &  & 0.130$^O$ & Magnetoabsorption (5K)  &     \cite{tanaka}  \\
                      &    &   &  0.117$^O$ &  High Field Magnetoabsorption   &\cite{galkowski} \\
                      E&    &   &  0.119$^C$ & T and $\rho$-resolv. Optical Spect. (RT) & \cite{yangliu} \\ 
%                      &    &  0.35 & $^C$  & ARPES   & \cite{takashi}\\
                      &   &  0.30(15) &  $^C$ & LEED-ARPES (RT) (along X-M) & \cite{fengzhou}\\
                      &   &  0.25(8) &  $^C$ & LEED-ARPES (RT) (along X-R)  & \cite{fengzhou}\\
                      &   &  0.27(2) &  $^C$ & ARPES (RT) (around M)  & \cite{feiyu}\\
%                      &  &  0.24 & $^C$  & ARPES   & \cite{takashi}\\
                      \cline{2-6} &  0.26 &  0.25 & 0.127$^C$  & DFT with GW and SOC &   \cite{bokdam}\\ 
                      &  0.29  & 0.31 & 0.150$^C$ & DFT with HSE and SOC  &   \cite{bah} \\
                      T &  0.25 & 0.30 & 0.136$^C$& DFT with SOC  &   \cite{choljun} \\
%                      &  0.20 & 0.24 & 0.109$^C$& DFT (Virtual Crystal Approx.)  &  \cite{ungi} \\
%                      &  0.26 & 0.29 & 0.137$^C$& DFT  &   \cite{nael} \\
%                      &  0.19 & 0.23 & 0.104$^C$& DFT with SOC  &  \cite{nael} \\
%                      T &  0.36 & 0.27 & 0.154$^C$& DFT with U  &  \cite{welch} \\
%                      &  0.32 & 0.31 & 0.158$^C$& DFT with SOC  &   \cite{martinez} \\
%                      &  0.20 & 0.28 & 0.117$^C$& DFT with SOC  &   \cite{saxena} \\
                      &  0.28 & 0.28 & 0.140$^T$ & DFT with HSE and SOC  &   \cite{bah} \\
%                      &  0.27 & 0.31 & 0.144$^T$& DFT with SOC  &  \cite{mosconi} \\                 
%                      &  0.35 & 0.46 & & DFT with U (Less.e-)  &   C &  \cite{welch} \\
%                      &  0.25 & 0.50 & 0.167$^T$& DFT ($\Gamma$-M)  &   \cite{Berdiyorov} \\
%                      &  0.50 & 0.29 & 0.183$^T$& DFT ($\Gamma$-Z)  &    \cite{Berdiyorov} \\
                      \hline \hline
$\textrm{MAPbCl}_{3}$  &  0.30  &  0.32 & 0.155$^C$  & DFT with GW and SOC &  \cite{bokdam}\\
                      T &  0.36  & 0.34 & 0.175$^C$ & DFT with SOC &  \cite{bah} \\
%                      &  0.34 & 0.47 & 0.197$^C$ & DFT  &   \cite{ungi2} \\
%                      T &  0.54 & 0.49 & 0.257$^C$& DFT with U  &   \cite{welch} \\
%                      &  0.37 & 0.51 & 0.214$^C$ & DFT with U (Less.e-)  &    \cite{welch} \\
                      &  0.37 & 0.32 & 0.172$^C$& DFT with SOC &  \cite{saxena} \\
\hline\hline
$\textrm{CsPbBr}_{3}$ &    &   &  0.126$^O$ &  High Field Magnetoabsorption &\cite{zang} \\
                     &    &   &  0.180$^O$ &  Magneto-optics (1.6K) &\cite{yakovlev} \\
                    &    & 0.26(2)  &  $^O$ &  ARPES at 33.5eV (along $\Gamma$-M)&\cite{puppin} \\
                    &    & 0.24(1)  &  $^O$ &  ARPES at 32eV (along $\Gamma$-M) &\cite{sajedi} \\                  
                    E&    & 0.22(1)  &  $^O$ &  ARPES at 41eV (along $\Gamma$-M) &\cite{sajedi} \\
                    &    & 0.20(1)  &  $^O$ &  ARPES at 35eV (along $\Gamma$-M)&\cite{sajedi} \\
                    &    & 0.29  &  $^O$ &  ARPES (along $\Gamma$-M)  &\cite{Zhiyuan} \\
                    &    & 0.26  &  $^O$ &  ARPES (along X-M)  &\cite{Zhiyuan} \\                   
                    \cline{2-6} &   &  & 0.102$^O$ & DFT with SOC &  \cite{filip}  \\ 
                     T & 0.26  & 0.27 & 0.132$^{O/C}$ & DFT with HSE and SOC at 300K (Ave.) &  \cite{baiqing}  \\
                     & -  & 0.23 & $^{O}$ & DFT with GW and SOC  &  \cite{sajedi}  \\
 %                    & 0.07  & 0.07 & 0.034$^O$ & DFT with LDA and SOC &  \cite{ghaithan}  \\ 
%                     & 0.17  & 0.06 & 0.047$^O$ & DFT with PBE-GGA &  \cite{ghaithan}  \\
%                     & 0.07  & 0.06 & 0.032$^O$ & DFT with PBE-GGA and SOC &  \cite{ghaithan}  \\
%                     & 0.24  & 0.08 & 0.059$^O$ & DFT with EV-GGA  &  \cite{ghaithan}  \\
%                     & 0.08  & 0.08 & 0.038$^O$ & DFT with EV-GGA and SOC &  \cite{ghaithan}  \\
%                     T & 0.22  & 0.07 & 0.054$^O$ & DFT with PBEsol-GGA &  \cite{ghaithan}  \\
%T                     & 0.07  & 0.07 & 0.034$^O$ & DFT with PBEsol-GGA and SOC &  \cite{ghaithan}  \\
%                     & 0.25  & 0.08 & 0.054$^O$ & DFT with mBJ-GGA &  \cite{ghaithan}  \\
%                     & 0.08  & 0.08 & 0.041$^O$ & DFT with mBJ-GGA and SOC &  \cite{ghaithan}  \\
%                     & 0.23  & 0.08 & 0.058$^O$ & DFT with umBJ-GGA &  \cite{ghaithan}  \\
%                     & 0.07  & 0.07 & 0.036$^O$ & DFT with umBJ-GGA and SOC &  \cite{ghaithan}  \\
%                     &  & 0.17 & $^O$ & DFT with HSE and SOC &  \cite{puppin}  \\
%                     & 0.13 & 0.13 & 0.065$^O$ & DFT with HSE or GW &  \cite{becker}  \\
                      \hline\hline
$\textrm{FAPbI}_{3}$ &    &   &  0.090$^O$ &  High Field Magnetoabsorption  &\cite{galkowski} \\
                   E &    &   &  0.095$^T$ &  High Field Magnetoabsorption  &\cite{galkowski} \\
                    &    &   &  0.098$^C$ &  T and $\rho$-resolv. Optical Spect.(RT) &\cite{yangliu} \\
                    \cline{2-6} 
%                   &  0.02 & 0.10 & 0.017$^C$ & DFT with SOC&  \cite{zeeshan}  \\
%                    &  0.69 & 0.05 & 0.043$^T$ & DFT with SOC &  \cite{zeeshan}  \\
%                    &0.43 & 0.21 & 0.141$^C$&DFT-PBE  & \cite{sanjun} \\
                    &0.11 & 0.16 &  0.065$^C$&DFT-PBE with SOC & \cite{sanjun} \\
                   % &0.44 & 0.25 &  0.159$^C$& DFT with HSE& \cite{sanjun} \\
                    T &0.18 & 0.21 &  0.097$^C$& DFT with HSE and SOC & \cite{sanjun} \\
                    &0.22 & 0.27 &  0.121$^C$& DFT with GW and SOC & \cite{zeeshan2} \\
%                      &  0.32 & 0.32 & 0.160$^T$ & DFT with SOC  &  \cite{mosconi} \\
%                        &  0.13 & 0.38 & 0.097$^T$ & DFT (Avg. values)   &  \cite{Berdiyorov} \\ 
%$\textrm{MAPbI}_{2}$Cl$_1$ &  0.27  & 0.37  &   &  DFT with SOC & C &\cite{yuping} \\ 
%                      & 0.61  & 0.33  &   &  DFT with SOC &  C &\cite{danli} \\
%                      &  0.27 & 0.33 & & DFT PBE  &   C &  \cite{nael} \\
%                      &  0.13 & 0.24 & & DFT with SOC  &   C &  \cite{nael} \\
%                      &  0.32 & 0.26 & & DFT with U  &   C &  \cite{welch} \\
%                      &  0.66 & 0.56 & & DFT with U (Less.e-)  &   C &  \cite{welch} \\
%50I/50Cl                 &  0.27 & 0.34 & & DFT  &   C &  \cite{ungi2} \\
%8$\%$ Cl                      & 0.19  & 0.28  &   &  DFT with SOC &  C &\cite{quarti} \\
%$\textrm{MAPbBr}_{2}$Cl$_1$ &  0.30  & 0.62  &   &  DFT with SOC &  C &\cite{nael} \\ 
  \hline\hline
\end{tabular}
\caption{Experimental (E) and theoretical (T) literature reports of the reduced effective mass and bare effective masses of electron ($m^{*}_e$) and hole ($m^{*}_h$) for MAPbI$_3$, MAPbBr$_3$, MAPbCl$_3$, CsPbBr$_3$ and FAPbI$_3$. \label{table:mass1}}
\end{table*}

\clearpage
\section{Kane and Pollmann-Buttner Polaron Model}
Summary of equations from the exciton-polaron model proposed by Kane et al. \cite{kane}. These converge to the Pollmann and Buttner model when $E_{k\ min}^{B}=0$ \cite{kane,pollmann}.
\begin{align}
    E_{BE}&=-|E_{\infty}| G_{1}G_{2}, \\
    G_{n}&=1-\left( 1- \frac{\epsilon_{\infty}}{\epsilon_{0}} \right) F_{n} ,\\
    F_{1}&=\frac{2}{\pi} F_{d} ; \ F_{2}=  \frac{2}{\pi} (2F-F_{d}), \\
    F_{d}&=F+2 \left( \eta_{1}\frac{\partial F}{\partial \eta_{1}} + \eta_{2}\frac{\partial F}{\partial \eta_{2}} \right), \\
    F&= \int_{0}^{\infty} \left( E_{k \ min}^{A} + A_{11}^{-1}+ A_{22}^{-1} + E_{k \ min}^{B}\right) dx, \\
    \beta&=\beta_{\infty}G_{1}, \ \beta_{\infty}=\frac{e^{2}\mu}{\hbar^{2}\epsilon_{\infty}}, \\
    \eta_{n}&=2\frac{\hbar^{2}\beta^{2}_{\infty}G_{1}^{2}}{m_{n}\hbar\omega_{LO}}, 
     \ |E_{\infty}|=\frac{e^{4}\mu}{2\hbar^{2}\epsilon_{\infty}^{2}}
\end{align}
with:
\begin{align}
    E_{k \ min}^{A}&= \frac{2A_{1}A_{2}A_{12}-A_{11}A_{2}^{2}-A_{22}A_{1}^{2}
    }{4 \left( A_{11}A_{22} -A_{12}^{2}\right)}\\
    A_{nn}&= 1+ \frac{\hbar^2 k^2}{2m_{n}\hbar\omega_{LO}} \\
    A_{1}&=A_{2}=2(1-R_{00}) \\
    A_{12}&=-R_{00}\\
    R_{00} &=-\frac{1}{\left( 1+ x^{2} \right)^2}, \ x = \frac{k}{2\beta}
\end{align}
and,
\begin{align}
    E_{k \ min}^{B}&= \frac{2B_{1}B_{2}B_{12}-B_{11}B_{2}^{2}-B_{22}B_{1}^{2}
    }{4 \left( B_{11}B_{22} -B_{12}^{2}\right)}\\
    B_{nn}&=\frac{4}{3}\beta^2 \sigma_{n}^{2} x^2 \left( 1+\eta_{1}+\eta_{2} + \eta_{n}x^2 \right) \\
    B_{n}&=-4\beta \left( \sigma_{n}x^{2}R_{00} + \frac{1}{3}\sigma_{n}^{2}x^{2}\eta_{n} + \sigma_{1}\sigma_{2} \left( \eta_{\tilde{n}}\delta + x^2R_{00} \right) \right)\\
    B_{12}&=4 \beta^2 \sigma_{1}\sigma_{2} \left(  \delta - R_{00}x^{4} + (\eta_{1}+\eta_{2})\delta \right)\\
    \sigma_{n}&=\frac{A_{12}A_{\tilde{n}}-A_{\tilde{n}\tilde{n}}A_{n}}{2(A_{11}A_{22}-A_{12}^{2})}\\
    \delta&=1-\frac{1}{x\textrm{Tan}(x)}, \ \tilde{n}=\tilde{1}=2, \ \tilde{2}=1.
\end{align}

\bibliographystyle{aip}
\bibliography{Reference2}